\documentclass[journal]{IEEEtran}
\usepackage{graphicx}
\usepackage{booktabs}
\usepackage{subfigure}
\usepackage{overpic}
\usepackage{amsmath}
\usepackage{flushend}
\usepackage{amssymb}
\usepackage{multirow}
\usepackage{makecell}
\usepackage{listings}
\usepackage{algorithm}
\usepackage{fontawesome}
\usepackage{color, soul}
\usepackage[absolute,overlay]{textpos}                    

\usepackage{hyperref}
\hypersetup{
	colorlinks=true,
	linkcolor=blue,
	urlcolor=blue,
	citecolor=blue
}
\usepackage{colortbl}
\usepackage[dvipsnames]{xcolor}
\definecolor{bg}{HTML}{e0f1ff}

\begin{document}

\title{Dual-Branch State-Displacement Network for Sea Surface Temperature Super-Resolution}
\author{
Wankun Chen,  
Feng Gao, \emph{Member}, \emph{IEEE},
Yanhai Gan, 
Chuanzheng Gong, \\
Xun Gong, 
Junyu Dong, \emph{Member}, \emph{IEEE},
Qian Du, \emph{Fellow}, \emph{IEEE}
\thanks{This work was supported in part by the Natural Science Foundation of Shandong Province under Grant ZR2024MF020, in part by the Key Research and Development Program of Shandong Province under Grant 2025CXPT185, and in part by the Natural Science
Foundation of China under Grant 42406192. \textit{(Corresponding author: Feng Gao and Junyu Dong)}

Wankun Chen, Feng Gao, Yanhai Gan, Chuanzheng Gong, and Junyu Dong are with the State Key Laboratory of Physical Oceanography, and also with the Frontiers Science Center for Deep Ocean Multispheres and Earth System, Ocean University of China, Qingdao 266100, China.

Xun Gong is with the State Key Laboratory of Physical Oceanography, Institute of Oceanographic Instrumentation, Shandong Academy of Sciences, Qingdao 266100, China

Qian Du is with the Department of Electrical and Computer Engineering, Mississippi State University, Starkville, MS 39762 USA.}}

\markboth{Journal of Selected Topics in Applied Earth Observations and Remote Sensing}{Shell}

\maketitle

\begin{abstract}
Sea surface temperature (SST) is a critical indicator of global climate change, yet satellite-derived SST imagery often suffers from coarse spatial resolution, limiting the ability to capture fine-scale thermal structures such as ocean fronts. To address this, we propose a Dual-Branch State-Displacement Network (DBSD-Net) for SST super-resolution. DBSD-Net adopts a dual-branch architecture: a wavelet frequency branch that explicitly separates low and high-frequency components via discrete wavelet transform for targeted processing, and a VGGUNet branch that extracts multi-scale semantic features from a frozen pre-trained VGG backbone. Within the wavelet branch, we introduce a Structural State Space Module (SSSM) with a Gated Structure Refinement (GSR) unit to efficiently capture long-range dependencies and enhance structural integrity, and a Displacement Gate Module (DGM) that learns a displacement field for geometry-aware modulation of high-frequency details, thereby mitigating spatially varying degradation.  Experiments on multiple public SST datasets demonstrate that DBSD-Net outperforms existing state-of-the-art methods and exhibits greater robustness at larger upscaling factors.
\end{abstract}


\IEEEpeerreviewmaketitle

\section{Introduction}

\IEEEPARstart {O}{cean} is a fundamental component of the climate system and plays an indispensable role in it. Sea surface temperature (SST) is a key indicator of the long-term warming trend in the global climate system \cite{Prochaska23tgrs}. It is an important parameter for analyzing the exchange of energy, momentum, and moisture between the ocean and the atmosphere. It also greatly influences climate patterns \cite{wentz00science}. SST data mainly come from numerical models and satellite observations \cite{4778913}, and integrate various physical, chemical and biological parameters \cite{COURTOIS201760}. It has been widely applied in the studies of ocean circulation and the coupling of atmosphere-ocean \cite{meng23tgrs}.

However, a persistent challenge lies in the inherent trade-off between spatial resolution and temporal coverage. Due to various observational and instrumental constraints \cite{liu2025ijcv}, the acquired resolution often falls short of the stringent demands of practical applications. Consequently, most satellite-derived SST products are limited to moderate spatial resolutions—for example, microwave sensors typically offer a resolution of approximately 25 km\cite{10197640}. This limitation has spurred continuous research efforts to enhance the resolution of SST data. There is therefore a pressing need to develop multi-scale remote sensing data fusion and reconstruction methods that can improve the resolution and application value of ocean observations, ultimately supporting detailed, long-term ocean monitoring \cite{lloyd22tgrs}.

Image super-resolution (SR) reconstruction techniques offer a key solution to this problem \cite{yao2026jstars}, while preserving the spatial coverage of infrared and microwave satellite observations \cite{sr23review}. In the field of computer vision, numerous deep learning-based SR methods have been developed, including convolutional neural networks (CNNs) \cite{srcnn, yan22tmm}, channel-wise attention mechanisms \cite{zhang18eccv}, and Transformer architectures \cite{swinir}. These methods have become a widely adopted approach in data-driven Earth system science \cite{8113128}. Compared with numerical ocean models, such deep learning methods possess a distinct advantage: they can effectively learn high-order representations to bridge the gap between low-resolution inputs and high-resolution outputs.

Among current approaches, Transformer-based SR methods \cite{dos20vit, swinir, cyz23spl, lqg24tmm, hjf22grsl} and state space model-based SR methods \cite{mambair2024eccv, zhao2024jstars, yang2026jstars,wang2024ele} have demonstrated promising potential. As the scale of observational data continues to expand \cite{8113128}, traditional deep learning methods often encounter challenges when applied to large-scale geophysical problems \cite{9714397, Hutgrs}, such as significant computational overhead \cite{hdc23iccv, 9714397}, and difficulties in extracting sparse data features \cite{Richard}. Although many SR networks suitable for natural images and remote sensing images have been developed \cite{11187314}, there are only a few networks specifically designed for SST. Similarly, these general models do not fully consider the characteristics of RSI, especially its rich low-frequency structural information and complex high-frequency texture details. Uneven blurring and geometric distortion during the reconstruction process are also a prominent problem.

Existing methods face several key limitations when applied to SR of SST remote sensing data. First, modeling global information through self-attention incurs substantial computational cost, as it entails quadratic computational complexity \cite{vim2024arxiv,ali2023mdpi}. Moreover, most methods process all features uniformly in the spatial domain, lacking the ability to perceive different frequency components \cite{wang2024tgrs1}. {Second, the degradation process inevitably induces spatially non-uniform blur and geometric distortion, which most existing methods fail to effectively mitigate. Third, the majority of approaches adopt a unified architecture and rarely exploit the rich semantic information embedded in multi-scale features. Moreover, excessively deep networks tend to discard phase information during hierarchical feature extraction, overlooking the need for further feature refinement in high-dimensional space and thereby limiting the model's reconstruction capacity.
} 

To overcome these limitations, we propose a dual-branch super-resolution framework, termed the Dual-Branch State Displacement Network (DBSD-Net), for enhancing the spatial resolution of SST data. It integrates two complementary branches to tackle { spatially varying degradation} and limited structural coherence in coarse-resolution SST images. An iterative wavelet decomposition branch first decomposes input features into low-frequency and high-frequency subbands via the discrete wavelet transform (DWT). The low-frequency component is processed by a Structural State Space Module (SSSM), which captures long-range structural dependencies through a two-dimensional selective scanning mechanism. The high-frequency component is refined by a Displacement Gate Module (DGM), which employs a learnable displacement field to adaptively suppress non-uniform blur and geometric distortion. In parallel, a VGGUNet branch leverages a pre-trained VGG19 backbone {enhanced with} UNet++-style dense skip connections to extract multi-scale semantic features. {The outputs of the two branches are fused and then upsampled via pixel shuffle.} Training is guided by a Fourier loss function to improve frequency-domain convergence, enabling more accurate recovery of fine SST structures.

The contributions of this work are as follows:
\begin{itemize}
    \item A dual-branch framework is proposed that combines wavelet decomposition, the Mamba state space model, and a learnable displacement mechanism, addressing the lack of frequency-aware processing and spatially varying degradation in SST super-resolution.
    \item A DBSD-Net architecture is developed, which leverages the linear-complexity global receptive field of Mamba for long-range dependencies, a displacement gate module to suppress non-uniform degradation, a VGGUNet branch for multi-scale semantic priors, and {a focal frequency loss with dynamically normalized spectral weights to emphasize frequency components exhibiting large reconstruction errors}.
    \item Plug-and-play Structural State Space (SSSM) and Displacement Gate (DGM) modules are designed to strengthen long-range dependency modeling and spatially adaptive detail refinement.
    \item Extensive experiments on GHRSST, OISST, and HYCOM datasets demonstrate that DBSD-Net consistently surpasses state-of-the-art methods in both quantitative metrics and visual quality, confirming its effectiveness for high-fidelity SST reconstruction.
\end{itemize}

\section{Related Work}

\subsection{Natural Image Super-Resolution}
Natural image super-resolution algorithms can be categorized into three types: interpolation-based methods, reconstruction-based methods, and learning-based methods \cite{chazhi, chong, dl}. Among them, the generalization ability of interpolation- and reconstruction-based methods remains a challenge \cite{ibp, spare}, whereas learning-based methods have become dominant due to their capacity to automatically learn feature representations \cite{dlm}. SRCNN was the first to apply convolutional neural networks to SR, employing a three-layer architecture \cite{srcnn} and establishing the subsequent research paradigm. VDSR \cite{vdsr} introduced residual learning to address training difficulties. Later works replaced interpolation-based preprocessing with learnable upsampling layers, further mitigating noise issues.

Recent advances in super-resolution techniques have shown \cite{chen2023cvpr, ray2024cvpr, zhang2026atd,li2023iccv,elan} that Transformer-based architectures are gradually replacing convolutional neural network (CNN)-based methods. SwinIR \cite{swinir} leverages the Swin Transformer with window-based self-attention for image restoration tasks. SRFormer \cite{zhou23iccv} introduces permuted self-attention, which balances attention between the channel and spatial dimensions. Another approach \cite{yoo23wacv} combines the local features of CNNs with the long-range multi-scale dependencies of Transformers.

\begin{figure*}[]
  \centering
  \includegraphics[width=7in]{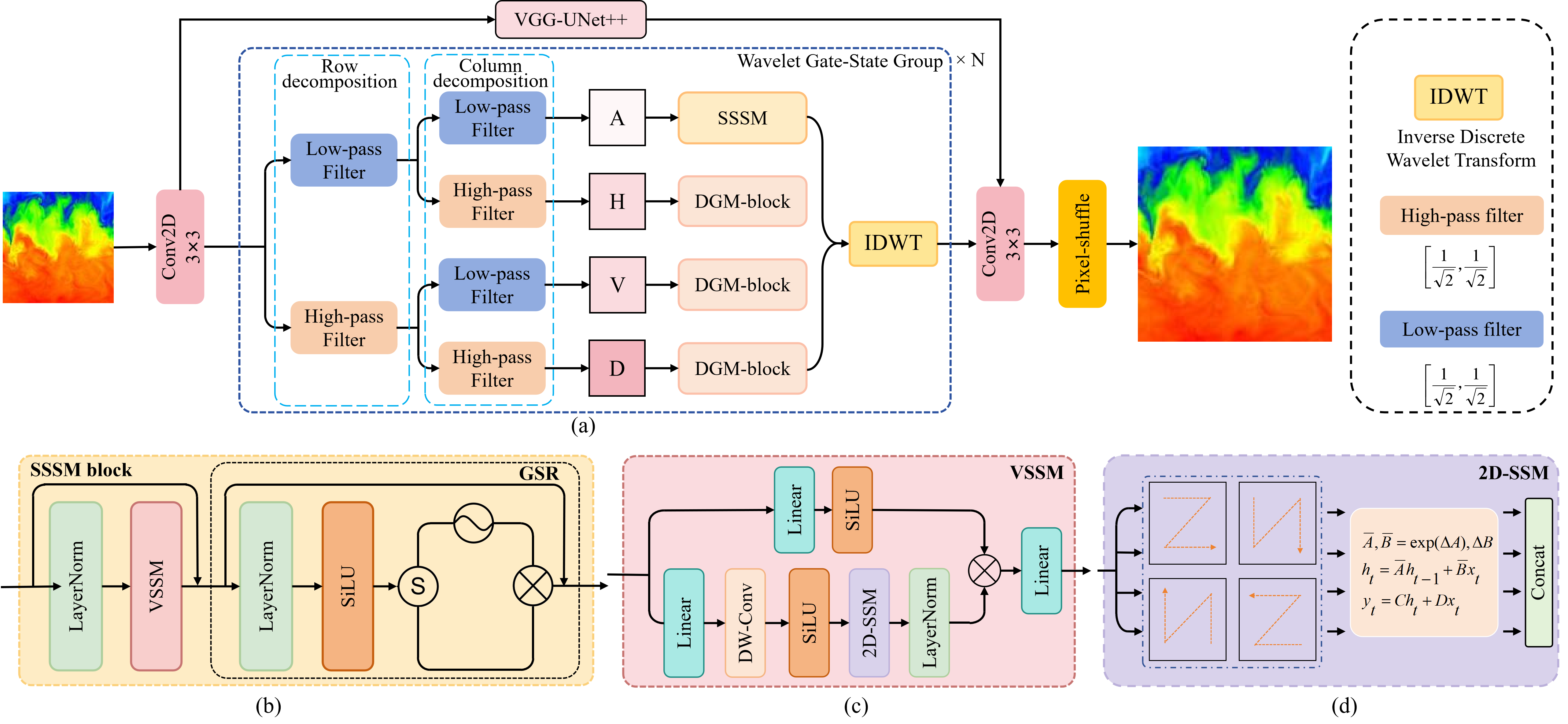}
  \caption{The overview of the Dual-Branch State-Displacement (DBSD) framework proposed for the SR of SST data is presented. (a) Illustrates the overall architecture of DBSD. DBSD consists of multiple stacked Wavelet Gate-state Groups (WGSG) branch and a VGGUNet++ branch. (b) Shows the structure of the SSSM block, which is composed of the VSSM block and the GSR module. (c) Displays the details of the VSSM block. (d) Shows the details of 2D-SSM.} 
  \label{network}
\end{figure*}

\subsection{SST Super-Resolution Based on Deep Learning}

Research on SST SR has been constrained by the inherently low resolution of observational data, yet deep learning approaches have attracted increasing attention in recent years. Ducournau et al. \cite{ducournau16prrs} first adapted SRCNN for this task. Su et al. \cite{su} combined a CNN with LightGBM and demonstrated that multi-sample training improved CNN performance. Ping et al. \cite{ping21jstars} designed an oceanic data reconstruction network with multi-scale feature extraction and multi-receptive field mapping, and found that appropriately deeper architectures yielded better results. Izumi et al. \cite{rs13183568} applied ESRGAN and compared it with classical methods, revealing the distinct behaviors of GAN-based and CNN-based architectures. Kim et al. \cite{kim23ija} introduced a GAN-driven spatio-temporal learning framework, and more recently, Zou et al. \cite{zou23rs} proposed a transformer-based model that integrates transformer and residual blocks for cross-scale feature learning.

Beyond the deep-learning super-resolution methods reviewed above, a largely separate line of research addresses SST reconstruction through physically motivated or statistical downscaling frameworks, including statistical downscaling, data assimilation, and multi-variable geophysical fusion \cite{wang2024rs,ding2024asr,wang2025sns,hoof2016jgra}. Although effective in their respective contexts, their task formulations differ fundamentally from the single-image super-resolution setting considered in this work, where only a low-resolution SST observation is provided as input. Consequently, numerical results reported under those settings are not directly comparable with ours.

\subsection{State Space Models}

State Space Models (SSMs) \cite{gu21ssm} have recently attracted considerable research interest due to their capability to model long-range dependencies while maintaining linear computational scalability with respect to sequence length. Mehta et al. \cite{mehta22long} incorporated gating mechanisms into SSM architectures, achieving enhanced performance in long-range language modeling. Mamba \cite{gu23mamba} introduced a selective scan mechanism, surpassing the Transformer paradigm in natural language modeling. VMamba \cite{liu2024nips} further extended this mechanism and applied it to visual tasks. A wealth of pioneering research has successfully applied Mamba to downstream tasks, such as remote sensing image classification \cite{wan2026jstars,shi2026jstars}, remote sensing image segmentation \cite{liu24cmunet,zhu2024jstars}, and image dehazing \cite{zhou24rsde}, all yielding encouraging results. Chen et al. \cite{chen2025mamba} introduced a wavelet-assisted Mamba super-resolution framework that leverages 2D SSMs for global temperature modeling and employs pixel difference convolution for texture refinement. Zhang et al. \cite{zhang2025jstars} proposed Algae-Mamba, a framework based on the visual state space model.

\section{Methodology}

The overall architecture of the proposed DBSD is illustrated in Fig. \ref{network}(a), which consists of four main stages: shallow feature extraction, dual-branch deep feature extraction, feature fusion, and image reconstruction.

Given an LR remote sensing image $\mathbf{I}_{LR}\in\mathbb{R}^{H\times W\times 3}$, a shallow feature is first extracted by a $3\times3$ convolution:
\begin{equation}
\mathbf{F}_0 = \operatorname{Conv}_{3\times3}(\mathbf{I}_{LR}),
\label{eq:shallow}
\end{equation}
where $\mathbf{F}_0\in\mathbb{R}^{H\times W\times C}$ denotes the initial feature map.

Subsequently, $\mathbf{F}_0$ is fed into two parallel deep feature extraction branches. The VGGUNet++ branch directly outputs multi‑scale semantic features $\mathbf{F}_{\text{vgg}}$. The feature branch is composed of $N$ cascaded Wavelet Gate-State Group
 (WGSG) modules, progressively extracting frequency‑aware features via
\begin{equation}
\mathbf{F}_i = \operatorname{WGSG}_i(\mathbf{F}_{i-1}) + \mathbf{F}_{i-1}, \quad i = 1,\dots,N,
\label{eq:dmwt_stack}
\end{equation}
yielding $\mathbf{F}_{\text{feat}} = \mathbf{F}_N$.

The outputs of the two branches are summed, passed through a $3\times3$ convolution, and added to the shallow feature $\mathbf{F}_0$ by a long skip connection:
\begin{equation}
\mathbf{F}_{\text{fused}} = \operatorname{Conv}_{3\times3}(\mathbf{F}_{\text{vgg}} + \mathbf{F}_{\text{feat}}) + \mathbf{F}_0.
\label{eq:fusion}
\end{equation}

Finally, a pixel‑shuffle layer upsamples the fused feature to the target resolution:
\begin{equation}
\mathbf{I}_{SR} = \operatorname{PixelShuffle}(\mathbf{F}_{\text{fused}}, s),
\label{eq:reconstruction}
\end{equation}
where $s$ is the upscaling factor.

\subsection{Wavelet Gate-State Group}
As the core building block of the feature branch, the Wavelet Gate State Group (WGSG) captures global structural dependencies and fine-grained local details in a frequency-aware manner. 
Using the Haar wavelet kernels $f_l = \frac{1}{\sqrt{2}}[1, 1]$ and $f_h = \frac{1}{\sqrt{2}}[1, -1]$, the input feature $\mathbf{F}_{\text{in}}$ is first decomposed, after which the low-frequency sub-band is processed by the SSSM for long-range coherence while the concatenated high-frequency sub-bands are refined by the DGM for geometry-aware modulation:
\begin{align}
\mathbf{X}_{LL}, \{\mathbf{X}_{HL}, \mathbf{X}_{LH}, \mathbf{X}_{HH}\} &= \operatorname{DWT}(\mathbf{F}_{\text{in}}),\label{eq:wgsg_dec}\\
\mathbf{Y}_{LL} &= \operatorname{SSSM}(\mathbf{X}_{LL}),\label{eq:wgsg_sssm}\\
\mathbf{Y}_{H} &= \operatorname{DGM}\bigl([\mathbf{X}_{HL}, \mathbf{X}_{LH}, \mathbf{X}_{HH}]\bigr).\label{eq:wgsg_dgm}
\end{align}

The enhanced sub-bands are then reconstructed by an inverse wavelet transform, and a residual connection is added to facilitate gradient propagation:
\begin{align}
\mathbf{F}_{\text{mid}} &= \operatorname{IWT}(\mathbf{Y}_{LL}, \mathbf{Y}_{H}),\label{eq:wgsg_iwt}\\
\mathbf{F}_{\text{out}} &= \mathbf{F}_{\text{mid}} + \mathbf{F}_{\text{in}}.\label{eq:wgsg_out}
\end{align}
This design enables the WGSG to simultaneously preserve global structural integrity and enhance local geometry-aware details.

\subsection{Structural State Space Module}

In SST super-resolution, capturing long-range dependencies is essential for reconstructing coherent global thermal structures, yet conventional CNNs are limited by local receptive fields, which motivates the design of the Structural State Space Module (SSSM), as illustrated in Fig. \ref{network}(b).

Given an input feature map $\mathbf{F}_l \in \mathbb{R}^{H\times W \times C}$, the module first applies layer normalization and feeds the result into a Vision State Space Module (VSSM) to establish global structural coherence. The output is subsequently refined by a Gated Structural Refinement (GSR) unit. The overall computation is formulated as:
\begin{equation}
    \mathbf{Z} = \mathrm{VSSM}\big(\mathrm{LN}(\mathbf{F}_l)\big) + \mathbf{F}_l,
    \label{eq:sssm1}
\end{equation}
\begin{equation}
    \mathbf{F}_{o} = \mathrm{GSR}\big(\mathbf{Z}\big) + \mathbf{Z},
    \label{eq:sssm2}
\end{equation}
where $\mathrm{LN}(\cdot)$ denotes layer normalization.

\textbf{Vision State Space Module (VSSM).} At the core of the VSSM lies a 2D state space model (2D-SSM), which builds upon the linear time-invariant formulation of State Space Models (SSMs)~\cite{gu21ssm}, and its structure is shown in Fig.~\ref{network}(c). A continuous-time SSM maps an input sequence $x(t)\in\mathbb{R}^N$ to an output $y(t)\in\mathbb{R}^N$ via an implicit latent state $h(t)\in\mathbb{R}^N$ through:
\begin{equation}
    \frac{d}{dt}h(t) = \mathbf{A}h(t) + \mathbf{B}x(t),\qquad
    y(t) = \mathbf{C}h(t) + \mathbf{D}x(t),
    \label{eq:ssm_continuous}
\end{equation}
where $\mathbf{A}\in\mathbb{R}^{N\times N}$ is the state transition matrix, $\mathbf{B}$ and $\mathbf{C}$ are projection matrices, and $\mathbf{D}$ is a skip connection.

To adapt to discrete data, zero-order hold discretization with timescale $\Delta$ yields $\overline{\mathbf{A}} = \exp(\Delta\mathbf{A})$ and $\overline{\mathbf{B}} = (\Delta\mathbf{A})^{-1}(\exp(\Delta\mathbf{A})-\mathbf{I})\,\Delta\mathbf{B}$, leading to the linear recurrence:
\begin{equation}
    h_t = \overline{\mathbf{A}} h_{t-1} + \overline{\mathbf{B}} x_t,\qquad
    y_t = \mathbf{C} h_t + \mathbf{D} x_t.
    \label{eq:ssm_discrete}
\end{equation}

The VSSM consists of two parallel branches and leverages the above discrete SSM to capture long-range dependencies with linear complexity. The discretization timescale $\Delta$ in Eq.~\eqref{eq:ssm_discrete} is learnable and input‑dependent, enabling the model to adaptively select relevant context.

Let $\mathbf{X} = \mathrm{LN}(\mathbf{F}_l)$ denote the input to the VSSM.

\textit{First branch:} A linear layer expands the channel dimension, followed by depth-wise convolution and SiLU activation to extract local structures, producing an intermediate feature $\mathbf{U}$.

The feature $\mathbf{U}$ then passes through a 2D-SSM and layer normalization to model global structural coherence. The 2D-SSM employs a four-way cross-scan strategy~\cite{liu2024nips} to efficiently cover the 2D space, as shown in Fig.~\ref{network}(d). Formally, let $\text{Scan}_k(\cdot)$ denote unfolding the feature map along the $k$-th traversal order ($k=1,\dots,4$), and $\text{Merge}(\cdot)$ the corresponding inverse operation. The 2D-SSM is computed as:
\begin{equation}
    \mathbf{v}_k = \text{SSM}\big(\text{Scan}_k(\mathbf{U})\big), \quad
    \mathbf{V} = \text{Merge}\big(\mathbf{v}_1, \mathbf{v}_2, \mathbf{v}_3, \mathbf{v}_4\big),
    \label{eq:2dssm}
\end{equation}
where $\text{SSM}(\cdot)$ denotes the discrete SSM defined in Eq.~\eqref{eq:ssm_discrete}. 

For brevity, we formally define the overall four‑way scanning and merging as $\mathrm{2D\text{-}SSM}(\mathbf{U}) \triangleq \mathrm{Merge}\big(\{\mathrm{SSM}(\mathrm{Scan}_k(\mathbf{U}))\}_{k=1}^4\big)$. This process captures global structural continuity across the entire feature map.

\textit{Second branch:} It simply applies a linear layer and SiLU activation. The outputs of the two branches are fused via element-wise multiplication, and a final linear projection restores the original channel dimension. The entire VSSM computation is summarized as:
\begin{equation}
    \mathbf{X}_1 = \mathrm{LN}\big(\mathrm{V}\big) = \mathrm{LN}\big(\mathrm{2D\text{-}SSM}(\mathbf{U})\big),
    \label{eq:vssm1}
\end{equation}
\begin{equation}
    \mathbf{X}_2 = \mathrm{SiLU}(\mathrm{FC}(\mathbf{X})),
    \label{eq:vssm2}
\end{equation}
\begin{equation}
    \mathrm{VSSM}(\mathbf{X}) = \mathrm{FC}(\mathbf{X}_1 \odot \mathbf{X}_2),
    \label{eq:vssm3}
\end{equation}
where $\mathbf{U} = \mathrm{SiLU}(\mathrm{DWConv}(\mathrm{FC}(\mathbf{X})))$, $\mathrm{FC}(\cdot)$ denotes a linear projection, $\mathrm{DWConv}(\cdot)$ is depth-wise convolution, and $\odot$ represents element-wise multiplication.

\textbf{Gated Structural Refinement (GSR).} The GSR unit further refines the structural features $\mathbf{Z}$ through a sequence of layer normalization, depth-wise convolution, and nonlinear gating. It first normalizes $\mathbf{Z}$ and processes it with depth-wise convolution; the result is then split along the channel dimension into two halves $\mathbf{Z}_1,\mathbf{Z}_2 \in \mathbb{R}^{H\times W\times C/2}$. The output of GSR is computed by a gating mechanism:
\begin{equation}
    \mathbf{Z}_o = \sigma(\mathbf{Z}_1) \odot \mathbf{Z}_2,
    \label{eq:gsr}
\end{equation}
where $\sigma(\cdot)$ is the sigmoid function. Thus, $\mathrm{GSR}(\mathbf{Z}) = \mathbf{Z}_o$, and the final refined feature in Eq.~\eqref{eq:sssm2} becomes $\mathbf{F}_o = \mathbf{Z}_o + \mathbf{Z}$. This gating operation enables individual channels to adaptively emphasize fine structural details, complementing the globally coherent structure established by the VSSM. 

\subsection{Displacement Gate Module}

\begin{figure}
    \centering
    \includegraphics[width=3.5in]{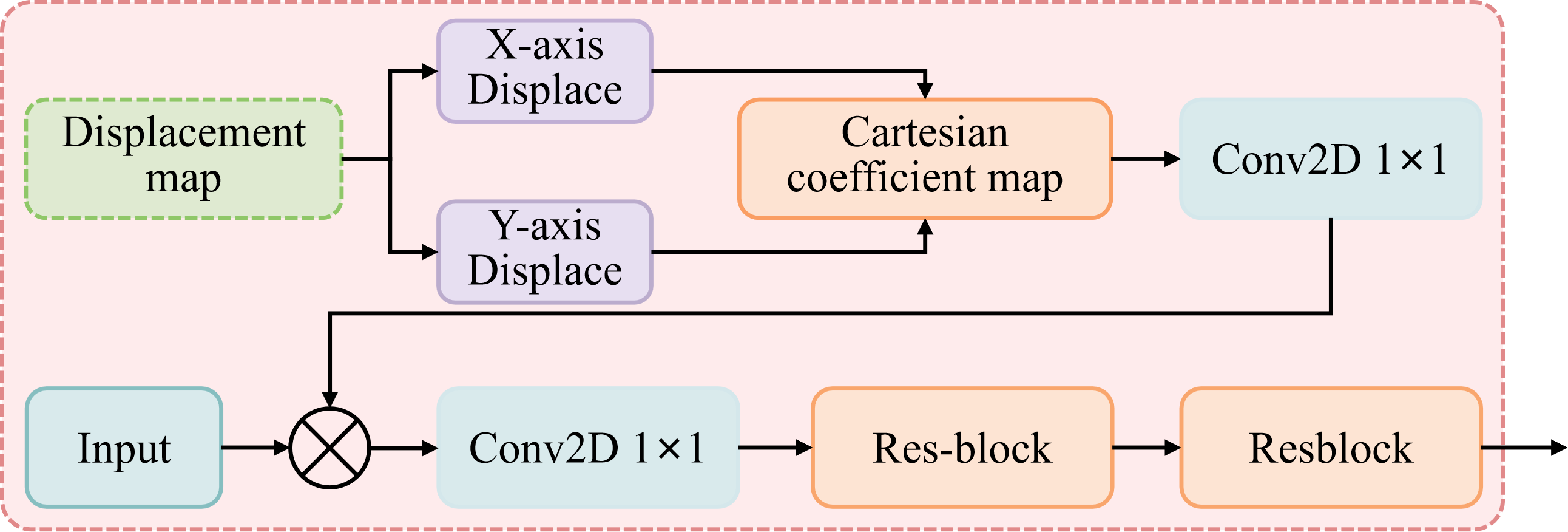}
    \caption{Details of the Displacement Gate Module (DGM). The DGM learns a compact displacement field, maps it to bilinear interpolation coefficients, and generates channel‑wise gating signals for geometry‑aware modulation of high‑frequency sub‑bands.}
    \label{fig_dgm}
\end{figure}

In remote sensing SST data, degradation often exhibits spatial variation including non‑uniform blur and geometric distortion, which standard CNNs cannot capture efficiently \cite{Howard_2023_BMVC}. This motivates a geometry‑aware design that explicitly encodes coordinate information for spatially adaptive feature modulation. To this end, we propose the Displacement Gate Module (DGM), whose structure is illustrated in Fig.~\ref{fig_dgm}.

First, DGM operates through a lightweight gating mechanism guided by a learned displacement field. Let $\mathbf{X} \in \mathbb{R}^{C \times H \times W}$ be the input feature map. A compact learnable tensor $\mathbf{D} \in \mathbb{R}^{2 \times h \times w}$ ($h \ll H, w \ll W$) stores sub‑pixel displacements $(\Delta x, \Delta y)$ on a coarse grid, which is bilinearly upsampled to full resolution to produce dense geometric context descriptors. {In our implementation, $h=w=16$, yielding $\mathbf{D}\in\mathbb{R}^{2\times16\times16}$, where the two channels represent the horizontal and vertical displacement components, respectively. The tensor $\mathbf{D}$ is directly initialized as a trainable model parameter rather than being dynamically predicted from each input image. It is jointly optimized with the other network parameters through backpropagation and shared across all samples processed by the corresponding DGM, while different DGM instances maintain their own independent displacement tensors.} The upsampling operation is formulated as

\begin{equation}
\mathbf{D}_{\mathrm{up}}
=
\operatorname{Upsample}_{\mathrm{bilinear}}
\left(\mathbf{D};H,W\right), ~~
\mathbf{D}_{\mathrm{up}}
\in\mathbb{R}^{2\times H\times W}.
\end{equation}

{Thus, the input feature map determines only the target resolution of the interpolated displacement field, rather than its values. Once training is completed, $\mathbf{D}$ remains fixed during inference and represents a module-specific, sample-shared spatial prior rather than an image-dependent displacement estimate.}

Second, per-pixel displacements are converted into a channel-wise gating map through a coefficient mapping and a $1{\times}1$ convolution.
{Specifically, a function $\psi$ bilinearly expands each $(d_x,d_y)$ into a $3{\times}3$ kernel: let $a=|d_x|$ and $b=|d_y|$.
The four non-zero weights $ab,\; a(1{-}b),\; (1{-}a)b,\; (1{-}a)(1{-}b)$ are placed at the corresponding adjacent pixels according to the signs of $(d_x,d_y)$, and the remaining five entries are zero. Consequently, the nine coefficients at each location sum to~$1$ without explicit normalisation.}
Stacking these kernels yields a coefficient volume, from which the gating map is obtained via a $1{\times}1$ convolution and ReLU~$\sigma$:
\[
\mathbf{G} = \sigma\bigl( \operatorname{Conv}_{1\times1}( \psi(\mathbf{D}_{\mathrm{up}}) ) \bigr) \in \mathbb{R}^{C\times H\times W}.
\]
{The unit-sum property holds only for the output of~$\psi$, while after the learnable projection, $\mathbf{G}$ is no longer constrained to sum to one, as it encodes channel-wise modulation strengths rather than interpolation weights. This learned mapping selectively amplifies the most relevant geometric cues per channel, achieving spatially adaptive modulation. Because $\mathbf{D}$ is shared across samples, the modulation varies spatially but is independent of individual input content.}

Then, the input features are modulated and channel‑mixed in one combined step: the features are first gated by $\mathbf{G}$ via element‑wise product, and a second $1\times1$ convolution immediately mixes the modulated channels to increase representational capacity:

\begin{equation}
\mathbf{X}_{\mathrm{mixed}}
=
\operatorname{Conv}_{1\times1}
\left(
\mathbf{X}\odot\mathbf{G}
\right).
\end{equation}

{It should be emphasized that neither $\mathbf{D}_{\mathrm{up}}$ nor its coefficient maps are used to resample, warp, or deform the input feature map. Instead, the displacement tensor parameterizes a position-dependent spatial--channel gate that modulates the feature responses. Therefore, DGM should be interpreted as a displacement-parameterized feature-modulation mechanism rather than an explicit geometric deformation or distortion-correction module.}

{
DGM leverages a learnable displacement field to construct spatially adaptive gating coefficients. Rather than explicitly resampling or warping the input features, it converts the displacement field into nine bilinear coefficient maps over a local $3 \times 3$ neighborhood; these coefficients are subsequently projected into a spatial-channel gate that adaptively modulates the input features. Thus, the learned displacement field serves as a shared spatial prior for feature modulation, rather than performing geometric deformation or distortion correction directly.
}

Finally, two sequential residual blocks, each composed of two $3\times3$ convolutions with ReLU and a skip connection, refine the modulated representation and improve training stability, yielding the output
{
\begin{equation}
\mathbf{Y}
=
\operatorname{ResBlock}_{2}
\left(
\operatorname{ResBlock}_{1}
\left(
\mathbf{X}_{\mathrm{mixed}}
\right)
\right).
\end{equation}
}
{This compact and progressive design enables DGM to enhance spatially varying high-frequency details through a learned position-dependent prior, while also complementing the global structure modeling of SSSM.}

\subsection{VGGUNet++ Module}

\begin{figure}
    \centering
    \includegraphics[width=3.5in]{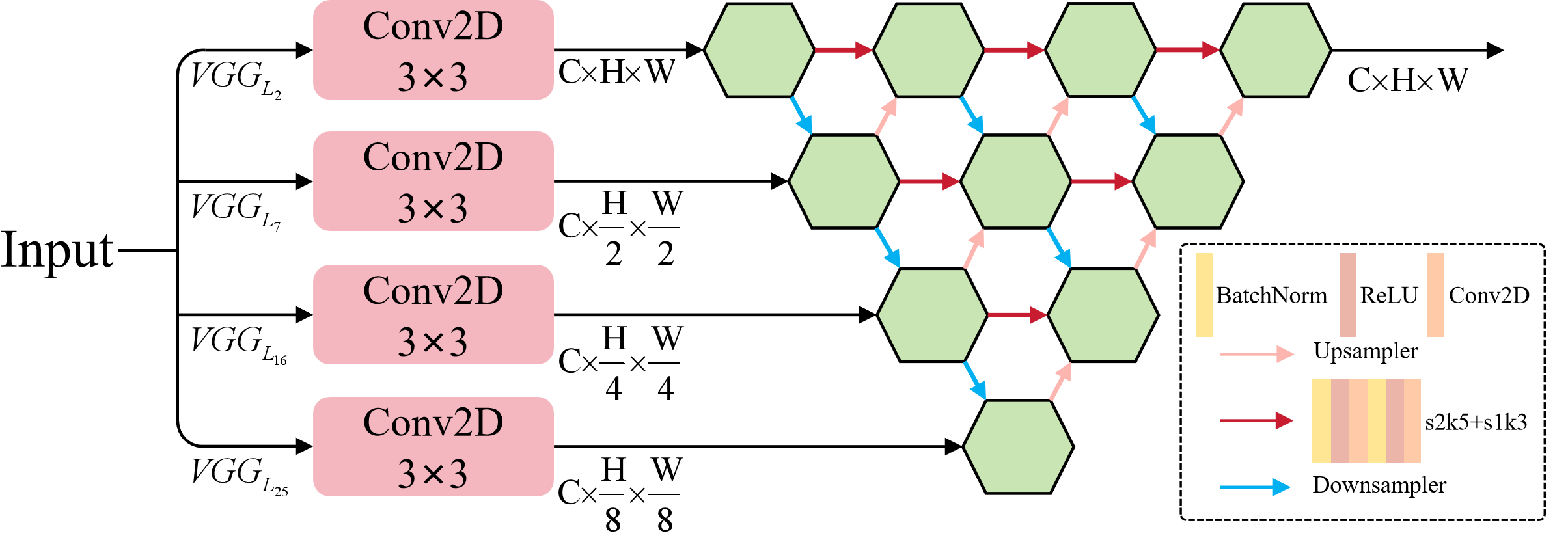}
    \caption{Details of the {VGGUNet++} Module. The {VGGUNet++} extracts features from four stages of a frozen VGG19, compresses them with grouped convolutions, and fuses them through dense nested skip connections with residual refinement for hierarchical multi‑scale semantic representation.}
    \label{VGGUNet}
\end{figure}

In remote sensing SST super-resolution, effectively capturing and fusing multi-scale features is crucial yet challenging. To address this, we propose the {VGGUNet++} module, which combines a pre-trained VGG19 backbone~\cite{vgg19} for hierarchical feature extraction with a UNet++~\cite{zhou2018unetpp} style nested fusion strategy; its architecture is illustrated in Fig.~\ref{VGGUNet}. This design establishes a {robust hierarchical representation that underpins the subsequent displacement gating and reconstruction stages}.

First, to construct multi-resolution feature maps, we extract outputs from four key layers of VGG19—denoted $L_2$, $L_7$, $L_{16}$, and $L_{25}$—which correspond to increasingly abstract semantic levels:
\begin{equation}
f_{k} = \text{VGG}_{L_k}(x), \quad k \in \{2,\,7,\,16,\,25\},
\label{eq:vgg_features}
\end{equation}
where $x$ is the input and $\text{VGG}_{L_k}$ represents the frozen output of the $k$‑th layer. Shallow layers ($L_2$, $L_7$) retain low‑level cues such as edges and textures, while deeper layers ($L_{16}$, $L_{25}$) capture high‑level semantic content.

Second,  to {unify channel dimensions while controlling computational cost}, each extracted feature map is compressed through a grouped convolution with $3\times 3$ kernels (16 groups), producing a set of lightweight multi‑scale representations with 48 channels:
\begin{equation}
y_{k} = \text{Conv}_{3\times 3}^{\text{(g=16)}}(f_{k}), \quad k \in \{2,\,7,\,16,\,25\}.
\label{eq:group_conv}
\end{equation}
{This grouping operation preserves spatial detail with lower parameter overhead than standard convolutions, while aligning the channel count across scales.}

Then, to effectively aggregate these semantically heterogeneous features, we adopt a {nested dense connectivity method inspired by UNet++}~\cite{zhou2018unetpp}. Unlike simple concatenation or summation, {iterative top‑down and lateral skip pathways are constructed to enable full-scale information exchange.}

At each fusion node, feature maps are combined via element‑wise addition after upsampling and subsequently refined by a lightweight residual block (denoted as $\mathcal{H}$). This summation‑based fusion maintains a constant channel dimension and reduces parameter overhead compared to concatenation, while still enabling dense multi‑scale information exchange.

{The fusion process for an encoder index $i$ (ordered from deep to shallow) and a fusion level $j$ is formalized as
\begin{equation}
x_{i,j} =
\begin{cases}
\mathcal{H}\!\bigl( y_{i} \bigr), & j = 0, \\[6pt]
\mathcal{H}\!\Bigl( \sum_{k=0}^{j-1} x_{i,k} \;+\; \mathcal{U}(x_{i+1,j-1}) \Bigr) \;+\; \mathcal{S}(y_{i}), & j > 0,
\end{cases}
\label{eq:nested_fusion}
\end{equation}
where {$y_{i}$} denotes the compressed feature from encoder depth $i$, $\mathcal{U}$ denotes bicubic upsampling (scale factor 2), and $\mathcal{S}$ represents a skip connection that directly passes the compressed feature.}

The $\mathcal{H}$ block implements a lightweight residual operation with channel‑wise feature recalibration. In our case, a four‑level encoder ($i=1,2,4,8$ corresponding to $L_2$, $L_7$, $L_{16}$, $L_{25}$) is used, and the fusion proceeds from the deepest level upwards until a single high‑resolution feature map is obtained. {To the best of our knowledge, VGGUNet++ represents the first application of UNet++‑style nested fusion to SST super‑resolution.}

Finally, the fused feature map captures both fine‑grained structural details and high‑level semantic context, providing a strong foundation for the subsequent displacement gate and reconstruction modules. 

{
\subsection{Focal Spectral Loss Function}
To enhance the reconstruction of fine-scale structures in SST super-resolution, we introduce a focal spectral loss\cite{fpl} that complements the pixel-wise spatial constraint. The base reconstruction loss is defined as the \(\ell_1\) difference between the super-resolved output \(I_{SR}\) and the high-resolution reference \(I_{HR}\):
\begin{equation}
    \mathcal{L}_{rec}=\left\|I_{SR}-I_{HR}\right\|_1 .
\end{equation}

To capture spectral discrepancies, the SST fields are transformed into the frequency domain via the two-dimensional Discrete Fourier Transform (DFT):
\begin{equation}
  \mathcal{F}(I)(u,v)
  =
  \frac{1}{HW}
  \sum_{x=0}^{H-1}
  \sum_{y=0}^{W-1}
  I(x,y)
  \exp\left[-2\pi i\left(\frac{ux}{H}+\frac{vy}{W}\right)\right],
\end{equation}
where \(H\) and \(W\) are the spatial dimensions of the field, \((x,y)\) and \((u,v)\) denote the spatial and frequency coordinates, and \(i\) is the imaginary unit. We write \(F_{SR}=\mathcal{F}(I_{SR})\) and \(F_{HR}=\mathcal{F}(I_{HR})\) for the corresponding complex spectra.

Rather than applying a fixed spectral weighting, we use a dynamic weight map that focuses on frequency components with large reconstruction errors. The magnitude discrepancy at each frequency is computed as
\begin{equation}
    \Delta(u,v)=\left|F_{SR}(u,v)-F_{HR}(u,v)\right|.
\end{equation}
A focal weight is then obtained by power-law normalization of this discrepancy:
\begin{equation}
    \omega(u,v)=
    \frac{\Delta(u,v)^{\alpha}}
    {\max_{u,v}\Delta(u,v)^{\alpha}+\epsilon},
\end{equation}
where \(\alpha\) controls the focusing strength and \(\epsilon\) ensures numerical stability. We set \(\alpha=1\) in all experiments. The focal spectral loss aggregates the weighted spectral differences:
\begin{equation}
    \mathcal{L}_{freq}
    =
    \frac{1}{HW}
    \sum_{u=0}^{H-1}
    \sum_{v=0}^{W-1}
    \omega(u,v)
    \left|F_{SR}(u,v)-F_{HR}(u,v)\right|^2 .
\end{equation}}

\section{Experimentals}
This section evaluates the proposed DBSD‑Net through comprehensive SR experiments conducted on three SST datasets. The performance of DBSD‑Net is compared against six state-of-the-art methods:  EDSR \cite{edsr2017cvpr}, CRAN \cite{cran2021iccv},  ELAN \cite{elan}, DAT \cite{dat2023iccv}, SAFM \cite{safmn2023iccv} and MambaIR \cite{mambair2024eccv}. Additionally, ablation studies are performed to assess the contribution of each key component within the proposed network.

\subsection{Dataset Settings}
The experiment utilized SST data from three distinct sources: the Hybrid Coordinate Ocean Model (HYCOM), the Optimal Interpolation Sea Surface Temperature (OISST) product, and the High Resolution Sea Surface Temperature Group (GHRSST) dataset. {The GHRSST benchmark, constructed from a single high-resolution SST image, is specifically designed to evaluate fine-scale spatial reconstruction by capturing rich spatial details from a fixed scene. Complementing this, the year-long HYCOM and OISST datasets encompass multiple dates and seasons, introducing substantial temporal and seasonal variations that enable the assessment of temporal robustness. For each dataset, a total of 1000 high-resolution images were obtained. The land and missing pixels are preprocessed by nearest neighbor filling, and no explicit masking of the land is performed. All reported metrics are computed on these filled fields, ensuring consistent tensor inputs across models but not isolating ocean-only regions. The LR input was generated by performing Bicubic downsampling on the HR reference field to obtain the corresponding low-resolution data, in order to ensure the pairing of real ground data and achieve a fair quantitative evaluation.} Subsequently, the dataset was divided into training and testing sets in a 4:1 ratio.

\textbf{HYCOM dataset} \cite{hycom}: HYCOM data were obtained from January to December 2016, with a spatial resolution of 1/12° and a temporal interval of 3 hours. The study area primarily covers unfrozen oceanic regions, including the North Pacific (NP, 5°N–45°N, 140°E–180°E), the Atlantic Gulf of Mexico (AGM, 5°S–45°S, 35°W–75°W), the North Indian Ocean (NIO, 5°N–35°S, 50°W–90°W), and the Equatorial Warm Pool (EWP, 20°N–20°S, 120°W–160°W).

\textbf{OISST dataset} \cite{oisst}: The OISST observational data used in this study were obtained from the National Oceanic and Atmospheric Administration (NOAA) website. The dataset has a spatial resolution of 1/4° and covers the period from January to December 2016.

\textbf{GHRSST dataset} \cite{ghrsst}: The Group for High Resolution Sea Surface Temperature (GHRSST) provides global, multi-sensor, high-resolution near-real-time SST products. The data used in this study were acquired on August 12, 2016, with a spatial resolution of 0.01°.

\subsection{Training Details and Evaluation Metrics}
The proposed model was trained using the PyTorch framework on an NVIDIA RTX 4080 GPU with 16 GB of memory. The Adam optimizer was employed with hyperparameters \(\beta_1 = 0.9\), \(\beta_2 = 0.999\), and \(\epsilon = 10^{-8}\). Training was conducted for 100 epochs with a batch size of 16 and a patch size of \(96 \times 96\). The initial learning rate was set to 0.0005 and reduced 50\% every 20 epochs. The activation function used throughout the network was RReLU. Super-resolution experiments were performed at scaling factors of 2, 3, and 4. 

{To evaluate the reconstruction performance of DBSD‑Net, six indicators were adopted: Peak Signal-to-Noise Ratio (PSNR), Structural Similarity Index Measure (SSIM), Gradient Magnitude Mean Absolute Error (Grad-MAE), Pearson correlation of gradient magnitudes (Grad-Corr), front-location F1-score (Front-F1), and front contour distance (FCD). PSNR and SSIM assess global pixel fidelity and structural similarity, while Grad-MAE and Grad-Corr quantify how accurately the intensity and sharpness are preserved, Front-F1 and FCD further measure the spatial alignment of detected fronts.

\begin{equation}
\mathrm{Grad\text{-}MAE}=
\frac{\sum_{i,j} B_{\mathrm{HR}}(i,j)\,
\bigl|g(\mathbf{T}_{\mathrm{SR}})_{i,j}-g(\mathbf{T}_{\mathrm{HR}})_{i,j}\bigr|}
{\sum_{i,j} B_{\mathrm{HR}}(i,j)},
\end{equation}

\begin{equation}
\begin{split}
&\mathrm{Grad\text{-}Corr}=\\
&\frac{\sum B_{\mathrm{HR}}(g_{\mathrm{SR}}-\bar{g}_{\mathrm{SR}})(g_{\mathrm{HR}}-\bar{g}_{\mathrm{HR}})}
{\sqrt{\sum B_{\mathrm{HR}}(g_{\mathrm{SR}}-\bar{g}_{\mathrm{SR}})^2\,
\sum B_{\mathrm{HR}}(g_{\mathrm{HR}}-\bar{g}_{\mathrm{HR}})^2}},
\end{split}
\end{equation}

\begin{align*}
P &= \frac{|\mathbf{F}_{\mathrm{SR}} \cap \mathrm{Dilate}(\mathbf{F}_{\mathrm{HR}}, r)|}{|\mathbf{F}_{\mathrm{SR}}|}, \\
R &= \frac{|\mathbf{F}_{\mathrm{HR}} \cap \mathrm{Dilate}(\mathbf{F}_{\mathrm{SR}}, r)|}{|\mathbf{F}_{\mathrm{HR}}|},
\end{align*}

\begin{equation}
\mathrm{Front\text{-}F1}=\frac{2PR}{P+R},
\end{equation}

\begin{equation}
\begin{split}
\mathrm{FCD}=
\frac{1}{|\partial\mathbf{F}_{\mathrm{HR}}|+|\partial\mathbf{F}_{\mathrm{SR}}|}
\Bigl(&\sum_{x\in\partial\mathbf{F}_{\mathrm{SR}}} d(x,\partial\mathbf{F}_{\mathrm{HR}})\\
+&\sum_{y\in\partial\mathbf{F}_{\mathrm{HR}}} d(y,\partial\mathbf{F}_{\mathrm{SR}})\Bigr),
\end{split}
\end{equation}
where $\partial\mathbf{F}$ denotes the set of front contour pixels and $d(\cdot,\cdot)$ is the Euclidean distance.}

{\subsection{Hyperparameter Sensitivity Analysis}
To determine the optimal configuration of DBSD, we analyzed different loss function configurations and pretrained weight freezing using the GHRSST dataset, and conducted sensitivity analyses on two key hyperparameters: the number of feature channels and the number of WGSG blocks.

\begin{figure} [!h]
  \centering
  \includegraphics[width=3.5in]{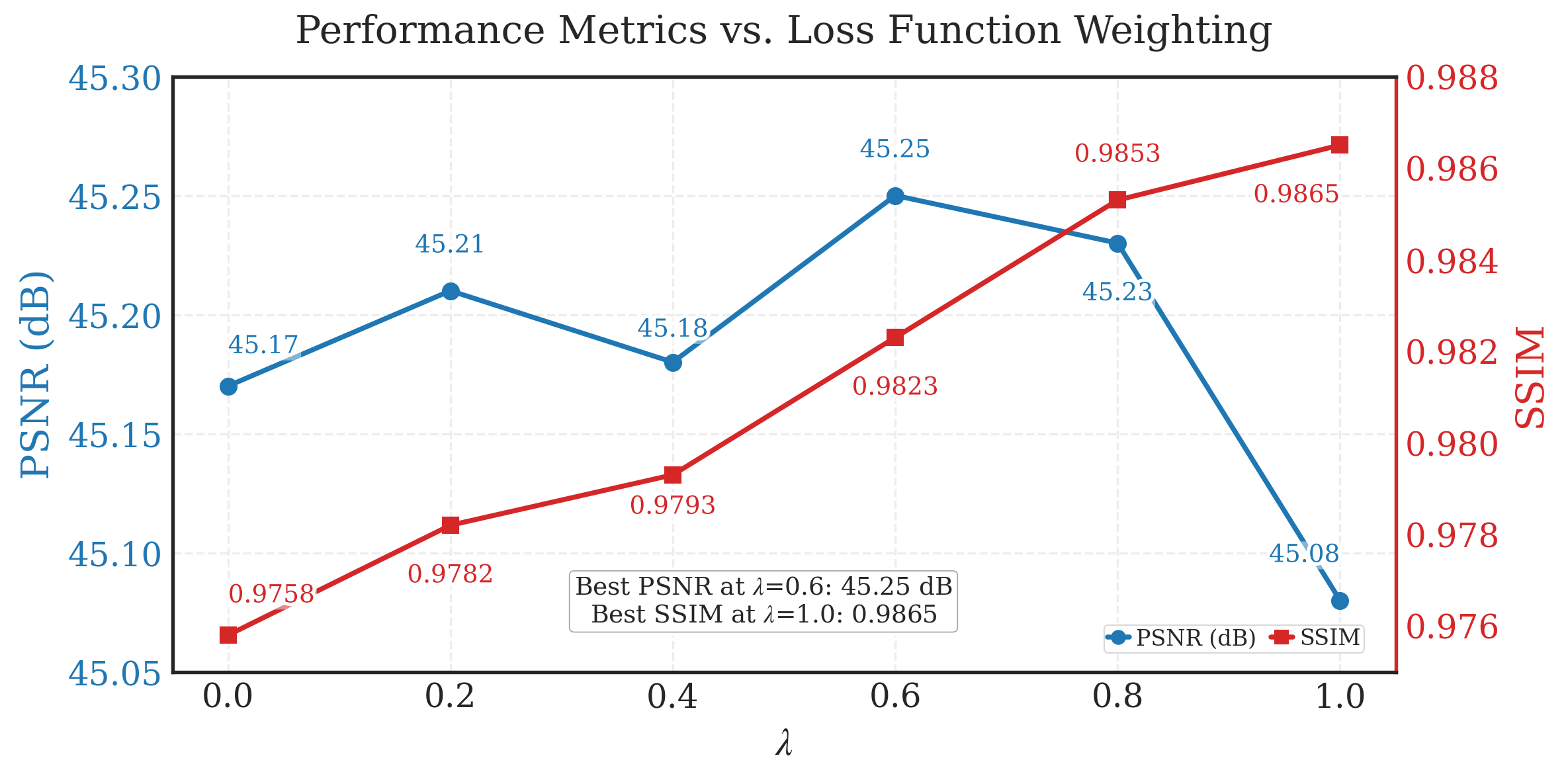}
  \caption{The influence of loss functions under different coefficients while training.} \label{fig:loss_weight_ablation}
\end{figure}

Firstly, to determine the optimal loss weights, we perform a grid search over the coefficients of the pixel-wise $\mathcal{L}_1$ loss and the focal frequency loss $\mathcal{L}_{\mathrm{FFL}}$, with the total loss defined as $\mathcal{L} = \lambda \mathcal{L}_{1} + (1-\lambda)\mathcal{L}_{\mathrm{FFL}}$. The experimental results are presented in Fig.~\ref{fig:loss_weight_ablation}. It is observed that the highest PSNR is attained at $\lambda=0.6$, while SSIM exhibits a monotonic increasing trend as $\lambda$ grows. Considering both metrics comprehensively, we select $\lambda=0.8$ for all subsequent experiments, as it achieves a favourable trade-off between pixel-level accuracy and frequency-domain detail preservation.

\begin{table}[!h]
\centering
{
\caption{Ablation study of VGG19 initialization and trainability on the GHRSST dataset.}
\label{tab:vgg_ablation}
\begin{tabular}{lcccc}
\toprule
Configuration & PSNR (dB) $\uparrow$ & SSIM $\uparrow$ & Params (M) \\
\midrule
PT-Frozen      & 46.57 & 0.9909 & 30.59    \\
Random-Frozen  & 46.53 & 0.9909 & 30.59      \\
PT-Trainable   & 46.52 & 0.9908 & 43.80 \\
Random-Trainable & 46.51 & 0.9907 & 43.80  \\
\bottomrule
\end{tabular}
}
\end{table}

Secondly, to further confirm the effectiveness of the VGG19 backbone, we conducted multiple ablation experiments to verify it. As shown in Table~\ref{tab:vgg_ablation}, the pretrained frozen VGG19 (PT-Frozen) achieves the best performance with a PSNR of 46.57\,dB and SSIM of 0.9909. Compared to the trainable variants, which introduce approximately 56\% more trainable parameters, PT-Frozen yields slightly higher accuracy while avoiding the additional optimization cost, demonstrating that fine-tuning the pretrained backbone brings no consistent benefit. Consequently, the pretrained frozen VGG19 is retained in the final model.}

\begin{table}[!h]
\centering
\caption{Impact of channel number on DBSD performance on GHRSST dataset.}
\scalebox{0.9}{
\begin{tabular}{ccccc}
\toprule
Number of Channels & PSNR $\uparrow$ & SSIM $\uparrow$ & FLOPs (G) & Params (M) \\ 
\midrule
32   & 43.14 & 0.9713 & 2.31    & 2.39  \\ 
48   & 44.39 & 0.9816 & 5.74   & 5.27  \\ 
64   & 45.23 & 0.9853 & 13.32  & 9.29  \\ 
96   & 46.84 & 0.9918 & 52.03  & 20.70 \\ 
128  & 47.35 & 0.9930 & 147.48 & 36.68 \\ 
144  & 47.39 & 0.9941 & 228.53 & 46.30 \\ 
\bottomrule
\end{tabular}}
\label{exp_channels}
\end{table}

{Thirdly, we further verified the impact of the number of channels on performance, all other parameters remained the same as in the previous experimental group, and we initially set 12 WGSG blocks.} As shown in Table~\ref{exp_channels}, increasing the channel dimension consistently improves reconstruction quality, with 128 channels achieving a PSNR of 47.35\,dB and SSIM of 0.9930. Further expanding to 144 channels yields a marginal PSNR gain of only 0.04\,dB while incurring a 48.4\% increase in FLOPs and a 26.4\% increase in parameters. {Hence, 128 channels is adopted as it offers the best accuracy–efficiency trade-off.}

\begin{table}[!h]
\centering
\caption{Impact of the number of WGSG blocks in the proposed DBSD on GHRSST dataset.}
\scalebox{0.9}{
\begin{tabular}{ccccc}
\toprule
Number of WGSG & PSNR $\uparrow$ & SSIM $\uparrow$ & FLOPs (G) & Params (M) \\
\midrule
 6  & 46.94 & 0.9922 & 74.29 & 18.42 \\
 8  & 46.76 & 0.9925 & 98.69 & 24.51 \\
10  & 47.05 & 0.9925 & 123.08  & 30.59 \\
12  & 46.87 & 0.9924 & 147.48  & 36.68 \\
16  & 46.86 & 0.9922 &  196.27 & 48.85 \\
\bottomrule
\end{tabular}}
\label{exp_numbers}
\end{table}

{Finally, regarding the performance analysis of the number of WGSG blocks,} Table~\ref{exp_numbers} shows that the best performance is attained with 10 blocks, achieving a PSNR of 47.05\,dB. Compared to the 16-block variant, this configuration significantly reduces both FLOPs and parameters, {while maintaining higher accuracy, demonstrating that  excessive stacking is detrimental to the convergence. Therefore, after comprehensively considering the balance between computational load and reconstruction quality, as well as the training difficulty of the model, the setting of 10 WGSG blocks was ultimately selected. In the subsequent experiments, DBSD maintained a configuration of 10 WGSG blocks, with each block having an input channel number set at 128.}
                        
\subsection{Computational Complexity and Runtime Efficiency}
\label{sec:efficiency}

To further analyze the computational efficiency of each method, we evaluated the practicability of DBSD and six representative SR methods, comparing the Parameters (Params), Floating Point Operations (FLOPs), latency, GPU peak memory, and throughput (images/second) of each method. After completing 30 warm-up iterations, 100 synchronous forward propagations were measured. Table~\ref{tab:efficiency_comparison} summarizes the present snapshot.

\begin{table}[!h]
\centering
\caption{Comparison of the super-resolution performance efficiency of different methods on simulated input, with input size of $[4,3,48,48]$, scaling factor of $\times4$, and averaging over 100 iterations.}
\label{tab:efficiency_comparison}
\setlength{\tabcolsep}{5pt}
\renewcommand{\arraystretch}{1.1}
\begin{tabular}{lccccc}
\toprule
Method & FLOPs & Params & Memory & Latency & Throughput \\
 & (G) & (M) & (MiB) & (ms) & (images/s) \\
\midrule
EDSR \cite{edsr2017cvpr}  & 18.30  & 1.37 & 90.5 & 1.55 & 1586.66 \\
CRAN \cite{cran2021iccv}  & 73.31 & 8.00 & 167.0 & 8.32 & 480.77 \\
ELAN \cite{elan}          & 29.57 & 1.43 & 138.5 & 22.13 & 180.75 \\
DAT \cite{dat2023iccv}    & 52.08 & 5.22 & 269.8 & 21.89 & 182.77 \\
SAFM \cite{safmn2023iccv} & 51.40 & 5.56 & 189.6 & 5.31 & 553.52 \\%
MambaIR \cite{mambair2024eccv} & 27.32 & 2.19  & 96.5 & 17.96 & 168.08 \\
\textbf{DBSD-Net} & 74.32 & 30.59 & 267.1 & 30.25 & 175.34 \\
\bottomrule
\end{tabular}
\end{table}

\begin{figure*}[t!]
  \centering
  \includegraphics[width=5in]{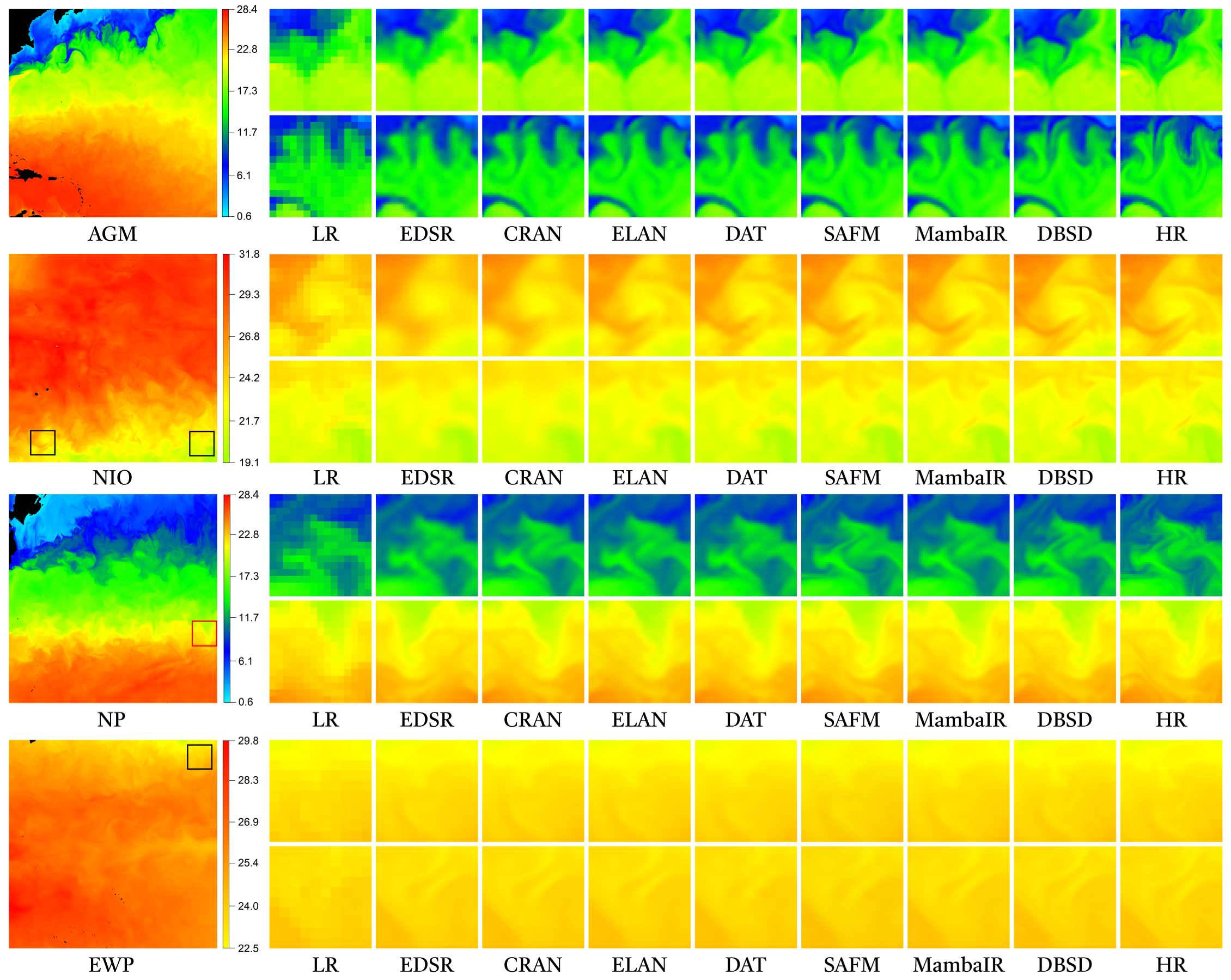}
  \caption{Qualitative analysis of different models on the HYCOM dataset with SR scale of 4, four representative regions selected from the validation dataset, each region intercepted a fixed size of data, and scaled to the same size, in order to best observe the details.} \label{fig:hycom}
\end{figure*}

\begin{table*}[!ht]
\footnotesize
\centering
\caption{Super-resolution performance of different methods on the HYCOM Dataset. PSNR and SSIM are calculated on the test set. \textcolor{red}{Red}: best, \textcolor{blue}{Blue}: second-best.}
\label{table_hycom}
\scalebox{0.9}{
\begin{tabular}{cc cccc cccc}
\toprule
\multirow{2}{*}{Method} & \multirow{2}{*}{SR scale} & \multicolumn{2}{c}{NP} & \multicolumn{2}{c}{AGM} & \multicolumn{2}{c}{EWP} & \multicolumn{2}{c}{NIO} \\
\cmidrule{3-10}
& & PSNR $\uparrow$ & SSIM $\uparrow$ & PSNR $\uparrow$ & SSIM $\uparrow$ & PSNR $\uparrow$ & SSIM $\uparrow$ & PSNR $\uparrow$ & SSIM $\uparrow$ \\
\midrule
\multirow{3}{*}{EDSR \cite{edsr2017cvpr}}
    & ×2 & 36.95 & 0.9837 & 45.63 & 0.9893 & 43.93 & 0.9928 & 37.64 & 0.9813 \\
    & ×3 & 31.34 & 0.9651 & 39.72 & 0.9812 & 37.25 & 0.9736 & 33.89 & 0.9685 \\
    & ×4 & 29.77 & 0.9341 & 38.57 & 0.9771 & 32.16 & 0.9332 & 32.31 & 0.9465 \\
\cmidrule{2-10}
\multirow{3}{*}{CRAN \cite{cran2021iccv}}
    & ×2 & 37.63 & 0.9871 & 45.96 & 0.9903 & 44.43 & \textcolor{blue}{0.9950} & 37.92 & 0.9853 \\
    & ×3 & 31.34 & 0.9651 & 39.72 & 0.9812 & 37.25 & 0.9736 & 33.89 & 0.9685 \\
    & ×4 & 30.42 & 0.9362 & 39.16 & 0.9786 & 36.80 & 0.9776 & 33.02 & 0.9389 \\
\cmidrule{2-10}
\multirow{3}{*}{ELAN \cite{elan}}
    & ×2 & 43.54 & \textcolor{blue}{0.9915} & \textcolor{blue}{49.62} & \textcolor{blue}{0.9930} & \textcolor{blue}{49.03} & 0.9946 & 44.07 & 0.9869 \\
    & ×3 & 37.70 & 0.9779 & 47.32 & 0.9893 & 45.86 & 0.9920 & 39.79 & 0.9764 \\
    & ×4 & \textcolor{blue}{35.73} & 0.9653 & \textcolor{blue}{45.58} & 0.9852 & \textcolor{blue}{44.10} & \textcolor{blue}{0.9894} & 37.96 & 0.9650 \\
\cmidrule{2-10}
\multirow{3}{*}{DAT \cite{dat2023iccv}}
    & ×2 & 38.19 & 0.9876 & 47.43 & 0.9920 & 44.91 & 0.9918 & 40.81 & 0.9893 \\
    & ×3 & 34.81 & 0.9709 & 44.89 & 0.9879 & 39.80 & 0.9837 & 38.23 & 0.9763 \\
    & ×4 & 32.54 & 0.9483 & 42.94 & 0.9823 & 41.31 & 0.9852 & 36.21 & 0.9562 \\
\cmidrule{2-10}
\multirow{3}{*}{SAFM \cite{safmn2023iccv}}
    & ×2 & \textcolor{blue}{43.67} & 0.9905 & 48.31 & 0.9918 & 47.71 & 0.9931 & \textcolor{blue}{44.81} & 0.9894 \\
    & ×3 & \textcolor{blue}{39.08} & \textcolor{blue}{0.9810} & 47.28 & 0.9894 & \textcolor{blue}{47.15} & 0.9921 & 40.31 & \textcolor{blue}{0.9799} \\
    & ×4 & 35.13 & 0.9627 & 43.61 & 0.9837 & 43.38 & 0.9885 & 37.09 & 0.9608 \\
\cmidrule{2-10}
\multirow{3}{*}{MambaIR \cite{mambair2024eccv}}
    & ×2 & 42.01 & \textcolor{blue}{0.9915} & 49.42 & 0.9929 & 48.53 & 0.9944 & 43.75 & \textcolor{blue}{0.9902} \\
    & ×3 & 37.72 & \textcolor{blue}{0.9810} & \textcolor{blue}{47.57} & \textcolor{red}{0.9899} & 46.36 & \textcolor{blue}{0.9924} & \textcolor{red}{41.52} & 0.9713 \\
    & ×4 & 35.35 & \textcolor{blue}{0.9673} & 44.86 & \textcolor{blue}{0.9853} & 42.75 & 0.9866 & \textcolor{blue}{37.99} & \textcolor{red}{0.9684} \\
\cmidrule{2-10}
\multirow{3}{*}{DBSD-Net}
    & ×2 & \textcolor{red}{43.78} & \textcolor{red}{0.9930} & \textcolor{red}{49.80} & \textcolor{red}{0.9936} & \textcolor{red}{50.02} & \textcolor{red}{0.9953} & \textcolor{red}{45.07} & \textcolor{red}{0.9932} \\
    & ×3 & \textcolor{red}{39.80} & \textcolor{red}{0.9875} & \textcolor{red}{47.58} & \textcolor{blue}{0.9897} & \textcolor{red}{47.29} & \textcolor{red}{0.9935} & \textcolor{blue}{41.49} & \textcolor{red}{0.9895} \\
    & ×4 & \textcolor{red}{36.36} & \textcolor{red}{0.9695} & \textcolor{red}{45.92} & \textcolor{red}{0.9861} & \textcolor{red}{44.90} & \textcolor{red}{0.9903} & \textcolor{red}{38.35} & \textcolor{blue}{0.9673} \\
\bottomrule
\end{tabular}}
\end{table*}

Table~\ref{tab:efficiency_comparison} presents a comprehensive efficiency comparison across representative methods. EDSR stands out as the most lightweight and fastest baseline, while SAFM also achieves a favorable balance with low latency and high throughput. Compared with these methods, DBSD maintains a moderate parameter budget, which is substantially leaner than that of CRAN, DAT, and SAFM, reflecting efficient parameter utilization. Its higher FLOPs  primarily originate from the fine-grained feature extraction mechanism tailored for accurate SST reconstruction, yet the resulting throughput remains comparable to that of ELAN and DAT. Although DBSD exhibits increased memory usage and latency, these costs are practically acceptable given the significant reconstruction quality gains it provides.

\subsection{Compare with the state of the art methods}

To evaluate the SR performance of our proposed DBSD-Net, we compared it against seven state-of-the-art methods: EDSR \cite{edsr2017cvpr}, CRAN \cite{cran2021iccv}, ELAN \cite{elan}, DAT \cite{dat2023iccv}, SAFM \cite{safmn2023iccv} and MambaIR \cite{mambair2024eccv}. As shown in Table~\ref{table_hycom}, DBSD-Net consistently delivers the best or near-best performance across all magnification scales and oceanic sub-regions. At $\times2$, it achieves the highest PSNR and SSIM in every region, demonstrating excellent restoration accuracy and structural fidelity. As the scale increases, its advantage remains clear: the performance drop from ×2 to ×4 is small, and in the few cases where it is slightly outperformed, the difference from the best value is negligible. Among the competing methods, MambaIR performs strongly in certain high‑scale settings, while ELAN and SAFM also show competitive results in several sub‑regions, yet none matches the comprehensive superiority and stability of DBSD-Net across all scales and regions. These results confirm that DBSD-Net provides accurate, consistent, and spatially adaptive SR for multi-scale ocean remote sensing data.

\begin{figure*}[t!]
  \centering
  \includegraphics[width=5in]{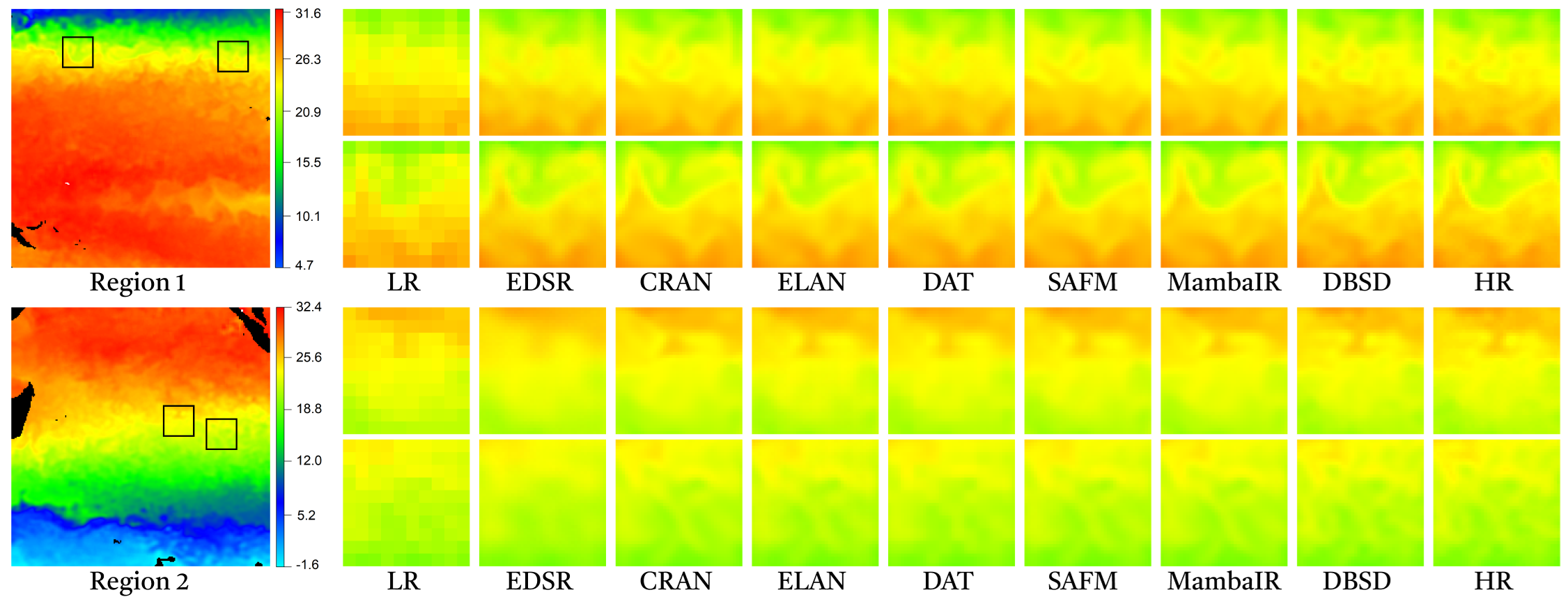}
  \caption{Qualitative analysis of different models on the OISST dataset with SR scale of 4, randomly selected two regions in the test dataset, each region intercepted a fixed size of data, and scaled to the same size in order to best observe the details.} \label{fig_oisst_res}
\end{figure*}

\begin{table*}[!ht]
\centering
\caption{Super-resolution performance of different methods on the OISST Dataset. PSNR and SSIM are calculated on the test set. \textcolor{red}{Red}: best, \textcolor{blue}{Blue}: second-best.}
\label{oisst}
\scalebox{0.9}{
\begin{tabular}{ccccccccc} 
\toprule
Scale & & EDSR \cite{edsr2017cvpr} & CRAN \cite{cran2021iccv} & ELAN \cite{elan} & DAT \cite{dat2023iccv} & SAFM \cite{safmn2023iccv} & MambaIR \cite{mambair2024eccv} & DBSD-Net \\
\midrule
\multirow{2}{*}{$\times$2} & PSNR$\uparrow$ & 41.83 & 42.14 & \textcolor{blue}{46.88} & 42.74 & 45.18 & 46.31 & \textcolor{red}{48.86} \\
& SSIM$\uparrow$ & 0.9887 & 0.9905 & \textcolor{blue}{0.9956} & 0.9913 & 0.9933 & 0.9943 & \textcolor{red}{0.9959} \\
\cmidrule{3-9}
\multirow{2}{*}{$\times$3} & PSNR$\uparrow$ & 36.36 & 37.42 & 38.80 & 37.72 & \textcolor{blue}{39.59} & 39.02 & \textcolor{red}{39.66} \\
& SSIM$\uparrow$ & 0.9757 & 0.9793 & \textcolor{blue}{0.9869} & 0.9806 & 0.9865 & 0.9862 & \textcolor{red}{0.9896} \\
\cmidrule{3-9}
\multirow{2}{*}{$\times$4} & PSNR$\uparrow$ & 34.84 & 35.11 & 35.28 & 33.15 & 35.66 & \textcolor{blue}{35.87} & \textcolor{red}{36.08} \\
& SSIM$\uparrow$ & 0.9566 & 0.9578 & 0.9679 & 0.9568 & 0.9669 & \textcolor{blue}{0.9705} & \textcolor{red}{0.9731} \\
\bottomrule
\end{tabular}}
\end{table*}

Fig.~\ref{fig:hycom} demonstrates DBSD-Net's superior reconstruction of fine-scale SST features across diverse ocean regions. In AGM and NP, where strong gradients and mesoscale patterns dominate, DBSD-Net preserves current boundaries and filamentary structures, whereas LR inputs show block artifacts and competing methods tend to over-smooth. In NIO, thermal transitions and local textures are well recovered. In EWP, differences are subtle due to smooth SST, but DBSD-Net still matches HR texture most closely. Overall, DBSD-Net excels at detail restoration and edge preservation, with reconstruction difficulty concentrated in regions of sharp fronts and dynamic current structures.

\textbf{Results on the OISST dataset.}
As shown in Table~\ref{oisst}, DBSD-Net consistently achieves the best PSNR and SSIM at all magnification factors, demonstrating excellent intensity fidelity and structural preservation. Its performance degrades only slightly as the scale increases from $\times2$ to $\times4$, indicating strong robustness to scale variation. Competing methods such as ELAN, SAFM, and MambaIR exhibit competitive results at individual scales, {but a clear performance gap relative to DBSD-Net persists. These results confirm that DBSD-Net can provide accurate and stable SR for SST data across multiple magnifications.}

Fig.~\ref{fig_oisst_res} illustrates that DBSD-Net achieves visually faithful SST super-resolution in both regions, with reconstructed patterns closely approaching the HR references. In Region 1, DBSD-Net better preserves the warm–cold transition and the local texture within the highlighted areas, while the LR input shows clear block artifacts and weaker structural details. In Region 2, where stronger spatial gradients and coastal influences are present, DBSD-Net maintains smoother yet sharper SST boundaries and recovers more consistent fine-scale variations than most comparison methods.

\begin{figure*}[!h]
  \centering
  \includegraphics[width=5in]{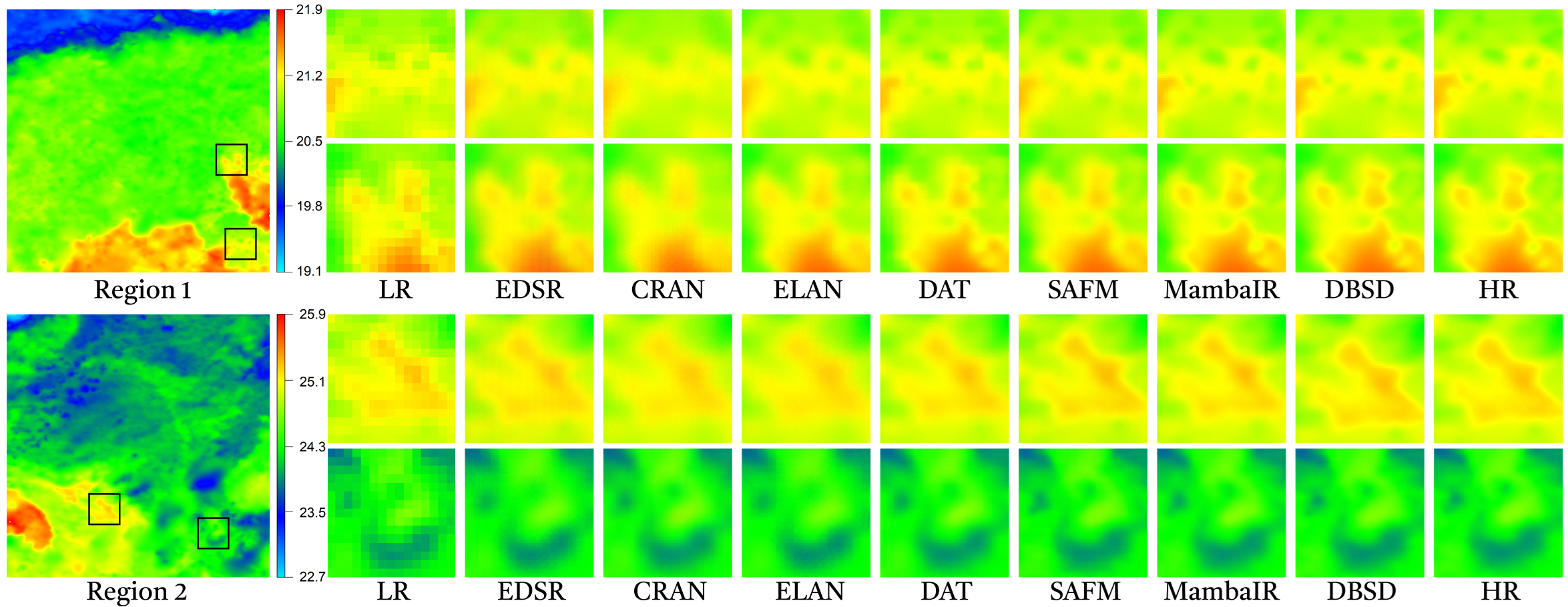}
  \caption{Qualitative analysis of different models on the GHRSST dataset with SR scale of 4, randomly selected two regions in the test dataset, each region intercepted a fixed size of data, and scaled to the same size in order to best observe the details.} \label{fig-ghrsst}
\end{figure*}

\begin{table*}[!ht]
\footnotesize
\centering
\caption{Super-resolution performance of different methods on the GHRSST Dataset. PSNR and SSIM are calculated on the test set. \textcolor{red}{Red}: best, \textcolor{blue}{Blue}: second-best.}
\label{tab-ghrsst}
\scalebox{0.9}{
\begin{tabular}{ccccccccc}  
\toprule
Scale & & EDSR \cite{edsr2017cvpr} & CRAN \cite{cran2021iccv} & ELAN \cite{elan} & DAT \cite{dat2023iccv} & SAFM \cite{safmn2023iccv} & MambaIR \cite{mambair2024eccv} & DBSD-Net \\
\midrule
\multirow{2}{*}{$\times$2} & PSNR$\uparrow$ & 49.36 & 48.37 & 51.57 & 51.13 & 50.20 & \textcolor{blue}{52.42} & \textcolor{red}{54.05} \\
& SSIM$\uparrow$ & 0.9911 & 0.9960 & 0.9971 & \textcolor{blue}{0.9973} & 0.9942 & \textcolor{blue}{0.9973} & \textcolor{red}{0.9982} \\
\cmidrule{3-9}
\multirow{2}{*}{$\times$3} & PSNR$\uparrow$ & 45.46 & 46.22 & \textcolor{blue}{49.87} & 48.97 & 48.42 & 48.56 & \textcolor{red}{50.66} \\
& SSIM$\uparrow$ & 0.9941 & 0.9926 & \textcolor{blue}{0.9958} & \textcolor{blue}{0.9958} & 0.9934 & 0.9955 & \textcolor{red}{0.9967} \\
\cmidrule{3-9}
\multirow{2}{*}{$\times$4} & PSNR$\uparrow$ & 42.99 & 44.21 & 46.79 & 46.06 & 45.11 & \textcolor{blue}{47.46} & \textcolor{red}{47.92} \\
& SSIM$\uparrow$ & 0.9721 & 0.9901 & 0.9921 & 0.9909 & 0.9875 & \textcolor{blue}{0.9931} & \textcolor{red}{0.9937} \\
\bottomrule
\end{tabular}}
\end{table*}

\textbf{Results on the GHRSST dataset.}
As shown in Table \ref{tab-ghrsst}, DBSD-Net consistently delivers the best PSNR across all magnifications and obtains the highest SSIM at $\times2$ and $\times3$, demonstrating outstanding intensity fidelity and structural preservation at common scales. At $\times4$, it still attains the top PSNR while its SSIM is slightly lower than those of MambaIR and ELAN, indicating a mild trade‑off at extreme upsampling. Among the other methods, MambaIR shows notable structural recovery at $\times4$, and ELAN performs well at multiple scales, yet neither matches the comprehensive advantage of DBSD-Net over the full range of magnifications. Overall, DBSD-Net exhibits strong robustness to scale variation, consistently maintaining leading or highly competitive accuracy.

Fig. \ref{fig-ghrsst} compares the SR results of different methods on GHRSST data. In Region 1, the LR input shows obvious block artifacts and blurred thermal transitions, while several competing methods tend to oversmooth local structures. In contrast, DBSD reconstructs finer SST textures and produces boundary patterns that are closer to the HR reference. {In Region 2, where the SST field contains more irregular spatial variations, DBSD better preserves the shape of warm and cold water masses and delineates their boundaries more clearly, without introducing noticeable spurious artifacts. }Overall, the visual results indicate that DBSD achieves more faithful structural reconstruction and stronger edge preservation than the compared methods.


{\subsection{Fine-scale Evaluation}

To further assess the reconstruction of fine-scale thermal structures, we introduce a front-aware evaluation protocol that focuses on regions with strong spatial gradients. Specifically, the gradient magnitude of the HR reference is computed using the Sobel operator, and pixels with responses above a predefined percentile are identified as frontal regions, where both intensity and positional consistency between the SR result and HR reference are evaluated. Lower Grad-MAE and FCD, and higher Grad-Corr and Front-F1, indicate better frontal structure preservation.

\begin{table*}[!h]
\footnotesize
\centering
\setlength{\tabcolsep}{4pt}
\caption{The detail-aware reconstruction performance of different super-resolution models under the HYCOM dataset ($\times$4 upscaling). 
\textcolor{red}{Red}: best, \textcolor{blue}{Blue}: second-best.}
\label{tab:front_eval}
\scalebox{0.9}{
\begin{tabular}{ccccccccc} 
\toprule
Region & Metric & EDSR \cite{edsr2017cvpr} & CRAN \cite{cran2021iccv} & ELAN \cite{elan} & DAT \cite{dat2023iccv} & SAFM \cite{safmn2023iccv} & MambaIR \cite{mambair2024eccv} & DBSD-Net \\
\midrule
\multirow{4}{*}{EWP}
    & Grad-MAE$\downarrow$ & 7.8649 & 7.6997 & 7.6383 & 7.6207 & \textcolor{blue}{7.6137} & 7.6139 & \textcolor{red}{7.6095} \\
    & Grad-Corr$\uparrow$ & 0.5785 & 0.6033 & 0.6133 & 0.6190 & 0.6197 & \textcolor{blue}{0.6205} & \textcolor{red}{0.6212} \\
    & Front-F1$\uparrow$ & 0.5588 & 0.5793 & \textcolor{red}{0.5818} & 0.5784 & \textcolor{blue}{0.5798} & 0.5785 & 0.5794 \\
    & FCD$\downarrow$ & 1.5558 & \textcolor{red}{1.5387} & 1.5526 & 1.5536 & \textcolor{blue}{1.5408} & 1.5571 & 1.5555 \\
\cmidrule{2-9}
\multirow{4}{*}{AGM}
    & Grad-MAE$\downarrow$ & 49.1438 & 48.4802 & 48.1875 & 47.9735 & 47.9740 & \textcolor{blue}{47.9335} & \textcolor{red}{47.9028} \\
    & Grad-Corr$\uparrow$ & 0.6186 & 0.6410 & 0.6460 & 0.6511 & 0.6506 & \textcolor{blue}{0.6524} & \textcolor{red}{0.6532} \\
    & Front-F1$\uparrow$ & 0.6977 & 0.7070 & 0.7087 & 0.7097 & \textcolor{blue}{0.7098} & \textcolor{red}{0.7100} & \textcolor{red}{0.7100} \\
    & FCD$\downarrow$ & \textcolor{blue}{0.8396} & 0.8828 & 0.8566 & 0.8509 & 0.8482 & 0.8424 & \textcolor{red}{0.8393} \\
\cmidrule{2-9}
\multirow{4}{*}{NIO}
    & Grad-MAE$\downarrow$ & 9.5402 & 9.2223 & 9.1275 & 9.0716 & 9.0676 & \textcolor{blue}{9.0654} & \textcolor{red}{9.0576} \\
    & Grad-Corr$\uparrow$ & 0.6408 & 0.6869 & 0.6949 & 0.6985 & 0.6987 & \textcolor{blue}{0.6995} & \textcolor{red}{0.7000} \\
    & Front-F1$\uparrow$ & 0.6235 & 0.6364 & \textcolor{blue}{0.6375} & 0.6351 & 0.6364 & 0.6349 & \textcolor{red}{0.6379} \\
    & FCD$\downarrow$ & 0.9521 & 0.9540 & \textcolor{blue}{0.9529} & 0.9530 & \textcolor{red}{0.9440} & 0.9543 & 0.9542 \\
\cmidrule{2-9}
\multirow{4}{*}{NP}
    & Grad-MAE$\downarrow$ & 30.9437 & 30.4921 & 30.2582 & 30.1143 & 30.1030 & \textcolor{blue}{30.0803} & \textcolor{red}{30.0625} \\
    & Grad-Corr$\uparrow$ & 0.6352 & 0.6538 & 0.6581 & 0.6616 & 0.6617 & \textcolor{blue}{0.6629} & \textcolor{red}{0.6637} \\
    & Front-F1$\uparrow$ & 0.6413 & 0.6543 & 0.6561 & 0.6567 & \textcolor{red}{0.6569} & \textcolor{blue}{0.6568} & \textcolor{blue}{0.6568} \\
    & FCD$\downarrow$ & 0.8605 & 0.8592 & 0.8522 & 0.8530 & \textcolor{blue}{0.8521} & 0.8523 & \textcolor{red}{0.8503} \\
\bottomrule
\end{tabular}}
\end{table*}

\begin{table*}[!h]
\centering
\caption{Ablation study of the SSSM, DGM, and VGGUNet modules. The PSNR and SSIM are reported on GHRSST datasets.}
\label{table_ablation0}
\resizebox{0.8\textwidth}{!}{
\begin{tabular}{l c c c c c c c c c c c c}
\toprule
\multirow{2}{*}{Case} & \multirow{2}{*}{SSSM} & \multirow{2}{*}{DGM} & \multirow{2}{*}{VGGUNet++} & \multicolumn{2}{c}{NIO} & \multicolumn{2}{c}{EWP} & \multicolumn{2}{c}{NP} & \multicolumn{2}{c}{AGM} \\
\cmidrule{5-6} \cmidrule{7-8} \cmidrule{9-10} \cmidrule{11-12}
& & & & PSNR$\uparrow$ & SSIM$\uparrow$ & PSNR$\uparrow$ & SSIM$\uparrow$ & PSNR$\uparrow$ & SSIM$\uparrow$ & PSNR$\uparrow$ & SSIM$\uparrow$ \\
\midrule
Case 1 & $\checkmark$ & $\times$     & $\times$     & 43.21 & 0.9801 & 41.74 & 0.9846 & 36.27 & 0.9532 & 32.99 & 0.9470 \\
Case 2 & $\times$     & $\checkmark$ & $\times$     & 42.97 & 0.9808 & 41.40 & 0.9833 & 36.16 & 0.9523 & 32.91 & 0.9453 \\
Case 3 & $\times$     & $\times$     & $\checkmark$ & 42.78 & 0.9766 & 41.45 & 0.9813 & 35.92 & 0.9469 & 32.71 & 0.9372 \\
\cmidrule{1-12}
Case 4 & $\times$     & $\checkmark$ & $\checkmark$ & 43.53 & 0.9826 & 41.80 & 0.9853 & 36.39 & 0.9566 & 33.05 & 0.9490 \\
Case 5 & $\checkmark$ & $\times$     & $\checkmark$ & 43.53 & 0.9828 & 41.79 & 0.9854 & 36.41 & 0.9571 & 33.09 & 0.9497 \\
Case 6 & $\checkmark$ & $\checkmark$ & $\times$     & 43.75 & 0.9831 & 41.95 & 0.9858 & 36.58 & 0.9586 & 33.24 & 0.9513 \\
\cmidrule{1-12}
DBSD-Net & $\checkmark$ & $\checkmark$ & $\checkmark$ & 43.82 & 0.9839 & 41.98 & 0.9860 & 36.60 & 0.9588 & 33.29 & 0.9517 \\
\bottomrule
\end{tabular}
}
\end{table*}

As shown in Table~\ref{tab:front_eval}, DBSD-Net consistently achieves the best Grad-MAE and Grad-Corr across all four datasets at $\times4$ upscaling, demonstrating outstanding gradient-level fidelity. It also attains the highest Front-F1 on AGM and NIO, and places second on NP, indicating strong preservation of edge and front structures. In terms of perceptual distance, DBSD-Net yields the lowest FCD on AGM and NP, and remains competitive on EWP and NIO, where SAFM and CRAN deliver marginally better scores. Among the other methods, MambaIR frequently obtains the second-best gradient metrics, while SAFM exhibits notable strength in Front-F1 and FCD on several datasets. Nevertheless, DBSD-Net maintains a clear overall advantage, particularly in reconstructing high-frequency front details with minimal gradient distortion. Overall, DBSD-Net exhibits robust front-aware reconstruction capability at the challenging $\times4$ scale, consistently delivering superior gradient accuracy and leading or highly competitive front perception quality across diverse image contents.}

\subsection{Ablation Study}
A bidirectional ablation strategy is employed to evaluate the contributions of the SSSM, DGM, and VGGUNet++ branch, where each module is assessed in isolation and through its removal from the full DBSD‑Net. All variants were trained on the HYCOM dataset for a 4$\times$ SR task. To save time, each variant was trained for 20 epochs. The quantitative results are shown in Table~\ref{table_ablation0}.

\textbf{SSSM module.} The isolated SSSM achieves the highest single‑module PSNR on NIO and AGM, indicating that state‑space modeling alone already captures meaningful structural dependencies in texture‑rich scenes. Removing SSSM from DBSD‑Net (Case~4) causes PSNR drops of 0.29\,dB on NIO and 0.24\,dB on AGM, while losses are smaller on EWP and NP . This confirms that the SSSM’s long‑range spatial modeling, enabled by the 2D selective scan, is most critical for maintaining coherence across sharp thermal fronts and complex mesoscale patterns.

\textbf{DGM module.} Isolated DGM performs comparably to SSSM on AGM but falls behind on NIO, suggesting that geometry‑aware modulation alone is less effective for capturing large‑scale coherence. Removing DGM from DBSD‑Net (Case~5) leads to PSNR declines of 0.29\,dB on NIO, 0.20\,dB on AGM, and 0.19\,dB on both NP and EWP. The pronounced drop on NIO highlights the contribution of the learned displacement field and gating mechanism in adaptively enhancing high‑frequency wavelet sub‑bands, particularly in regions with strong gradients and fine details.

\textbf{VGGUNet++ module.} The isolated VGGUNet++ yields the lowest performance across all datasets, especially on NIO and AGM. Comparing Case~6 to DBSD‑Net, adding VGGUNet++ brings marginal PSNR gains of 0.07\,dB on NIO, 0.03\,dB on EWP, 0.02\,dB on NP, and 0.05\,dB on AGM. This indicates that the frozen multi‑scale semantic features supply a modest auxiliary prior, while the dominant reconstruction capability stems from the SSSM–DGM synergy in frequency‑decomposed feature refinement.

\section{Conclusion}

In this paper, we proposed DBSD‑Net, a dual‑branch framework for SST super‑resolution that synergistically combines frequency‑decomposed structural modeling with multi‑scale semantic priors. To jointly capture long‑range thermal coherence and fine‑scale local details, the core feature branch adopts stacked Wavelet Gate State Groups, where the SSSM processes low‑frequency subbands for global spatial dependencies while the DGM refines high‑frequency subbands with a learnable displacement field for geometry‑aware modulation, effectively suppressing spatially varying degradation. {A parallel VGGUNet++ branch further supplies hierarchical semantic context via a frozen VGG‑19 with nested UNet++ style fusion. Bidirectional ablation studies reveal that the SSSM–DGM synergy drives the primary performance gain, with the VGGUNet++ branch contributing a stable auxiliary enhancement. }Extensive experiments on multiple SST datasets confirm that DBSD‑Net consistently surpasses state‑of‑the‑art methods in accuracy and visual quality, validating its effectiveness and efficiency.

Building upon these findings,{ future work will extend the framework to realistic SST super-resolution by incorporating observation-based degradation models and cross-sensor benchmarks with strict spatiotemporal collocation. Mask-aware training and ocean-only evaluation will be adopted to handle land and missing pixels. We also plan to develop content-adaptive geometric priors for the displacement gate, explore lightweight state-space formulations for faster inference, and extend the frequency-aware design to other remote sensing restoration tasks.
}

\bibliography{source}

@ARTICLE{yao2026jstars,
  author={Yao, Yu and Wang, Hengbin and Gao, Xiang and Xing, Ziyao and Zhang, Xiaodong and Zhao, Yuanyuan and Li, Shaoming and Liu, Zhe},
  journal={IEEE Journal of Selected Topics in Applied Earth Observations and Remote Sensing}, 
  title={DTWSTSR: Dual-Tree Complex Wavelet and Swin Transformer Based Remote Sensing Images Super-Resolution Network}, 
  year={2026},
  volume={19},
  number={},
  pages={4730-4747},
  doi={10.1109/JSTARS.2026.3651075}}

@article{Prochaska23tgrs,
  author={Prochaska, J. Xavier and Guo, Erdong and Cornillon, Peter C. and Buckingham, Christian E.},
  journal={IEEE Transactions on Geoscience and Remote Sensing}, 
  title={The Fundamental Patterns of Sea Surface Temperature}, 
  year={2023},
  volume={61},
  pages={1-19}}

@article{liu2025ijcv,
  author  = {Liu, Yuanye and Dian, Renwei and Li, Shutao},
  title   = {{Low-Rank Transformer for High-Resolution Hyperspectral Computational Imaging}},
  journal = {International Journal of Computer Vision},
  year    = {2025},
  volume  = {133},
  number  = {2},
  pages   = {809--824},
  issn    = {1573-1405},
  doi     = {10.1007/s11263-024-02203-7},
  url     = {https://doi.org/10.1007/s11263-024-02203-7}
}

@InProceedings{chen2023cvpr,
    author    = {Chen, Xiangyu and Wang, Xintao and Zhou, Jiantao and Qiao, Yu and Dong, Chao},
    title     = {Activating More Pixels in Image Super-Resolution Transformer},
    booktitle = {Proceedings of the IEEE/CVF Conference on Computer Vision and Pattern Recognition (CVPR)},
    month     = {June},
    year      = {2023},
    pages     = {22367-22377}}

@InProceedings{ray2024cvpr,
    author    = {Ray, Abhisek and Kumar, Gaurav and Kolekar, Maheshkumar H},
    title     = {CFAT: Unleashing TriangularWindows for Image Super-resolution},
    booktitle = {Proceedings of the IEEE/CVF Conference on Computer Vision and Pattern Recognition (CVPR)},
    year      = {2024},
    pages     = {26120-26129}}

@INPROCEEDINGS{li2023iccv,
  author={Li, Xiang and Dong, Jiangxin and Tang, Jinhui and Pan, Jinshan},
  booktitle={2023 IEEE/CVF International Conference on Computer Vision (ICCV)}, 
  title={DLGSANet: Lightweight Dynamic Local and Global Self-Attention Network for Image Super-Resolution}, 
  year={2023},
  volume={},
  number={},
  pages={12746-12755},
  doi={10.1109/ICCV51070.2023.01175}}

@article{zhang2026atd,
    title={ATD: Improved Transformer With Adaptive Token Dictionary for Image Restoration},
    author={Zhang, Leheng and Long, Wei and Li, Yawei and Zhou, Xingyu and Zhao, Xiaorui and Gu, Shuhang},
    journal={IEEE Transactions on Pattern Analysis and Machine Intelligence},
    year={2026},
    publisher={IEEE}
}

@ARTICLE{zhang2025jstars,
  author={Zhang, Yaoteng and Wang, Shuaipeng and Chen, Yanlong and Wei, Shiqing and Xu, Mingming and Liu, Shanwei},
  journal={IEEE Journal of Selected Topics in Applied Earth Observations and Remote Sensing}, 
  title={Algae-Mamba: A Spatially Variable Mamba for Algae Extraction From Remote Sensing Images}, 
  year={2025},
  volume={18},
  number={},
  pages={14324-14337},
  doi={10.1109/JSTARS.2025.3571988}}

@ARTICLE{wan2026jstars,
  author={Wan, Xiaoqing and Mo, Dongtao and He, Yupeng and Chen, Feng and Li, Zhize},
  journal={IEEE Journal of Selected Topics in Applied Earth Observations and Remote Sensing}, 
  title={MD2F-Mamba: Multidirectional Depthwise Convolution and Dual-Branch Mamba Feature Fusion Networks for Hyperspectral Image Classification}, 
  year={2026},
  volume={19},
  number={},
  pages={6214-6238},
  doi={10.1109/JSTARS.2026.3657648}}

@ARTICLE{zhu2024jstars,
  author={Zhu, Qinfeng and Fang, Yuan and Cai, Yuanzhi and Chen, Cheng and Fan, Lei},
  journal={IEEE Journal of Selected Topics in Applied Earth Observations and Remote Sensing}, 
  title={Rethinking Scanning Strategies With Vision Mamba in Semantic Segmentation of Remote Sensing Imagery: An Experimental Study}, 
  year={2024},
  volume={17},
  number={},
  pages={18223-18234},
  doi={10.1109/JSTARS.2024.3472296}}

@inproceedings{liu2024nips,
 author = {Liu, Yue and Tian, Yunjie and Zhao, Yuzhong and Yu, Hongtian and Xie, Lingxi and Wang, Yaowei and Ye, Qixiang and Jiao, Jianbin and Liu, Yunfan},
 booktitle = {Advances in Neural Information Processing Systems},
 doi = {10.52202/079017-3273},
 editor = {A. Globerson and L. Mackey and D. Belgrave and A. Fan and U. Paquet and J. Tomczak and C. Zhang},
 pages = {103031--103063},
 publisher = {Curran Associates, Inc.},
 title = {VMamba: Visual State Space Model},
 url = {https://proceedings.neurips.cc/paper_files/paper/2024/file/baa2da9ae4bfed26520bb61d259a3653-Paper-Conference.pdf},
 volume = {37},
 year = {2024}
}

@Article{wang2024rs,
  author  = {Wang, Shuo and Li, Xiaoyan and Zhu, Xueming and Li, Jiandong and Guo, Shaojing},
  title   = {Spatial Downscaling of Sea Surface Temperature Using Diffusion Model},
  journal = {Remote Sensing},
  volume  = {16},
  year    = {2024},
  number  = {20},
  pages   = {3843},
  doi     = {10.3390/rs16203843},
  issn    = {2072-4292},
}

@article{ding2024asr,
  author  = {Haiyong Ding and Xiaoyuan Qin},
  title   = {Downscaling study of microwave sea surface temperature products based on FY-3C satellite},
  journal = {Advances in Space Research},
  volume  = {74},
  number  = {5},
  pages   = {2117-2132},
  year    = {2024},
  doi     = {10.1016/j.asr.2024.05.074},
  issn    = {0273-1177},
}

@InProceedings{wang2025sns,
  author    = {Wang, Chen and Behrens, Erik and Ma, Hui and Chen, Gang and Huang, Victoria},
  editor    = {Gong, Mingming and Song, Yiliao and Koh, Yun Sing and Xiang, Wei and Wang, Derui},
  title     = {Climate Downscaling Monthly Coastal Sea Surface Temperature Using Convolutional Neural Network and Composite Loss},
  booktitle = {AI 2024: Advances in Artificial Intelligence},
  year      = {2025},
  publisher = {Springer Nature Singapore},
  address   = {Singapore},
  pages     = {303--315},
  isbn      = {978-981-96-0348-0}
}

@article{hoof2016jgra,
  author  = {Hoffmann, P. and Katzfey, J. J. and McGregor, J. L. and Thatcher, M.},
  title   = {Bias and variance correction of sea surface temperatures used for dynamical downscaling},
  journal = {Journal of Geophysical Research: Atmospheres},
  volume  = {121},
  number  = {21},
  pages   = {12,877-12,890},
  doi     = {10.1002/2016JD025383},
  year    = {2016}
}

@inproceedings{zhou2018unetpp,
  author    = {Zhou, Zongwei and Rahman Siddiquee, Md Mahfuzur and Tajbakhsh, Nima and Liang, Jianming},
  title     = {{UNet++}: A Nested U-Net Architecture for Medical Image Segmentation},
  booktitle = {Deep Learning in Medical Image Analysis and Multimodal Learning for Clinical Decision Support},
  year      = {2018},
  pages     = {3--11},
  publisher = {Springer International Publishing},
  address   = {Cham},
  isbn      = {978-3-030-00889-5}
}

@ARTICLE{shi2026jstars,
  author={Shi, Rui and Zhu, Xingyuan and Wu, Yongxiang and Gao, Hongmin and Dong, Jiaping and Wang, Gaoxu},
  journal={IEEE Journal of Selected Topics in Applied Earth Observations and Remote Sensing}, 
  title={MSFE-Mamba: Multiscale Frequency-Enhanced Mamba for Hyperspectral Image Classification}, 
  year={2026},
  volume={19},
  number={},
  pages={5234-5247},
  doi={10.1109/JSTARS.2026.3653204}}

@ARTICLE{chen2025mamba,
  author={Chen, Wankun and Gao, Feng and Gan, Yanhai and Cao, Jingchao and Dong, Junyu and Du, Qian},
  journal={IEEE Transactions on Geoscience and Remote Sensing}, 
  title={Wavelet-Assisted Mamba for Satellite-Derived Sea Surface Temperature Super-Resolution}, 
  year={2025},
  volume={63},
  number={},
  pages={1-12},
  doi={10.1109/TGRS.2025.3616324}}

@inproceedings{Howard_2023_BMVC,
author    = {Sunny Howard and Peter Norreys and Andreas Döpp},
title     = {CoordGate: Efficiently Computing Spatially-Varying Convolutions in Convolutional Neural Networks},
booktitle = {34th British Machine Vision Conference 2023, {BMVC} 2023, Aberdeen, UK, November 20-24, 2023},
publisher = {BMVA},
year      = {2023},
url       = {https://papers.bmvc2023.org/0744.pdf}
}

@ARTICLE{11187314,
  author={Chen, Wankun and Gao, Feng and Gan, Yanhai and Cao, Jingchao and Dong, Junyu and Du, Qian},
  journal={IEEE Transactions on Geoscience and Remote Sensing}, 
  title={Wavelet-Assisted Mamba for Satellite-Derived Sea Surface Temperature Super-Resolution}, 
  year={2025},
  volume={63},
  number={},
  pages={1-12},
  doi={10.1109/TGRS.2025.3616324}}

@inproceedings{dos20vit,
  title={An Image is Worth 16x16 Words: Transformers for Image Recognition at Scale},
  author={Dosovitskiy, Alexey and Beyer, Lucas and Kolesnikov, Alexander and Weissenborn, Dirk and Zhai, Xiaohua and Unterthiner, Thomas and  Dehghani, Mostafa and Minderer, Matthias and Heigold, Georg and Gelly, Sylvain and Uszkoreit, Jakob and Houlsby, Neil},
  booktitle={Proceedings of International Conference on Learning Representations (ICLR)},
  year={2021},
  pages={1-12}}

@article{lqg24tmm,
  author={Liu, Qingguo and Gao, Pan and Han, Kang and Liu, Ningzhong and Xiang, Wei},
  journal={IEEE Transactions on Multimedia}, 
  title={Degradation-Aware Self-Attention Based Transformer for Blind Image Super-Resolution}, 
  year={2024},
  volume={26},
  pages={7516-7528}}

@article{cyz23spl,
  author={Chen, Yuzhen and Wang, Gencheng and Chen, Rong},
  journal={IEEE Signal Processing Letters}, 
  title={Efficient Multi-Scale Cosine Attention Transformer for Image Super-Resolution}, 
  year={2023},
  volume={30},
  pages={1442-1446}}

@article{hjf22grsl,
  author={Hu, Jin-Fan and Huang, Ting-Zhu and Deng, Liang-Jian and Dou, Hong-Xia and Hong, Danfeng and Vivone, Gemine},
  journal={IEEE Geoscience and Remote Sensing Letters}, 
  title={Fusformer: A Transformer-Based Fusion Network for Hyperspectral Image Super-Resolution}, 
  year={2022},
  volume={19},
  pages={1-5}}

@InProceedings{hdc23iccv,
    author    = {Han, Dongchen and Pan, Xuran and Han, Yizeng and Song, Shiji and Huang, Gao},
    title     = {FLatten Transformer: Vision Transformer using Focused Linear Attention},
    booktitle = {Proceedings of the IEEE/CVF International Conference on Computer Vision (ICCV)},
    year      = {2023},
    pages     = {5961-5971}}

@InProceedings{cran2021iccv,
    author={Zhang, Yulun and Wei, Donglai and Qin, Can and Wang, Huan and Pfister, Hanspeter and Fu, Yun},
    booktitle={2021 IEEE/CVF International Conference on Computer Vision (ICCV)}, 
    title={Context Reasoning Attention Network for Image Super-Resolution}, 
    year={2021},
    pages={4258-4267},
    doi={10.1109/ICCV48922.2021.00424}}

@InProceedings{dat2023iccv,
    title={Dual Aggregation Transformer for Image Super-Resolution},    
    author={Zheng Chen and Yulun Zhang and Jinjin Gu and L. Kong and Xiaokang Yang and Fisher Yu},
    booktitle={2023 IEEE/CVF International Conference on Computer Vision (ICCV)},
    year={2023},
    pages={12278-12287}}

@InProceedings{safmn2023iccv,
    author={Sun, Long and Dong, Jiangxin and Tang, Jinhui and Pan, Jinshan},
    booktitle={2023 IEEE/CVF International Conference on Computer Vision (ICCV)}, 
    title={Spatially-Adaptive Feature Modulation for Efficient Image Super-Resolution}, 
    year={2023},
    volume={},
    number={},
    pages={13144-13153},
    doi={10.1109/ICCV51070.2023.01213}}

@InProceedings{edsr2017cvpr,
    title={Enhanced Deep Residual Networks for Single Image Super-Resolution},
    author={Bee Lim and Sanghyun Son and Heewon Kim and Seungjun Nah and Kyoung Mu Lee},
    booktitle={IEEE Conference on Computer Vision and Pattern Recognition Workshops (CVPR workshop)},
    year={2017},
    pages={1132-1140}}

@ARTICLE{zhao2024jstars,
  author={Zhao, Guangwei and Wu, Haitao and Luo, Dexiang and Ou, Xu and Zhang, Yu},
  journal={IEEE Journal of Selected Topics in Applied Earth Observations and Remote Sensing}, 
  title={Spatial–Spectral Interaction Super-Resolution CNN–Mamba Network for Fusion of Satellite Hyperspectral and Multispectral Image}, 
  year={2024},
  volume={17},
  number={},
  pages={18489-18501},
  doi={10.1109/JSTARS.2024.3469184}}

@ARTICLE{wang2024tgrs1,
  author={Wang, Sheng and Han, Boxun and Yang, Linzhe and Zhao, Chaoyue and Liang, Aokang and Hu, Chunying and Yang, Feng and Xu, Fu},
  journal={IEEE Transactions on Geoscience and Remote Sensing}, 
  title={Robust Remote Sensing Super-Resolution With Frequency Domain Decoupling for Multiscenarios}, 
  year={2024},
  volume={62},
  number={},
  pages={1-13},
  doi={10.1109/TGRS.2024.3406516}}

@Article{wang2024ele,
AUTHOR = {Wang, Wei and Zhao, Pei and Lei, Weimin and Ju, Yingjie},
TITLE = {ACMamba: A State Space Model-Based Approach for Multi-Weather Degraded Image Restoration},
JOURNAL = {Electronics},
VOLUME = {13},
YEAR = {2024},
NUMBER = {21},
ARTICLE-NUMBER = {4294},
URL = {https://www.mdpi.com/2079-9292/13/21/4294},
ISSN = {2079-9292},
DOI = {10.3390/electronics13214294}}

@ARTICLE{yang2026jstars,
  author={Yang, Jiawei and Ren, Hongliang},
  journal={IEEE Journal of Selected Topics in Applied Earth Observations and Remote Sensing}, 
  title={Rof-Mamba: Rotary-and-fourier-enhanced Mamba Network for Remote Sensing Image Super-Resolution}, 
  year={2026},
  volume={},
  number={},
  pages={1-21},
  doi={10.1109/JSTARS.2026.3691533}}

@InProceedings{mambair2024eccv,
    title={MambaIR: A Simple Baseline for Image Restoration with State-Space Model},
    author={Hang Guo and Jinmin Li and Tao Dai and Zhihao Ouyang and Xudong Ren and Shu-Tao Xia},
    booktitle={Proceedings of the European Conference on Computer Vision (ECCV)},
    year={2024},
    pages={ 222-241 }}

@article{wentz00science,
author = {Wentz, Frank J. and Gentenmann, Chelle and Smith, Deborah and Chelton, Dubley},
journal = {Science},
title = {Satellite Measurements of Sea Surface Temperature Through Clouds},
year = {2000},
volume = {288},
number = {5467},
pages = {847--850}}

@article{meng23tgrs,
author={Meng, Yuxin and Gao, Feng and Rigall, Eric and Dong, Ran and Dong, Junyu and Du, Qian},
journal={IEEE Transactions on Geoscience and Remote Sensing}, 
title={Physical Knowledge-Enhanced Deep Neural Network for Sea Surface Temperature Prediction}, 
year={2023},
volume={61},
pages={1-13}}

@article{10197640,
  author={Prochaska, J. Xavier and Guo, Erdong and Cornillon, Peter C. and Buckingham, Christian E.},
  journal={IEEE Transactions on Geoscience and Remote Sensing}, 
  title={The Fundamental Patterns of Sea Surface Temperature}, 
  year={2023},
  volume={61},
  pages={1-19}}

@ARTICLE{lloyd22tgrs,
  author={Lloyd, David T. and Abela, Aaron and Farrugia, Reuben A. and Galea, Anthony and Valentino, Gianluca},
  journal={IEEE Transactions on Geoscience and Remote Sensing}, 
  title={Optically Enhanced Super-Resolution of Sea Surface Temperature Using Deep Learning}, 
  year={2022},
  volume={60},
  pages={1-14}}

@inproceedings{4778913,
  author={Cornillon, Peter and Eichmann, Andrew},
  booktitle={IEEE International Geoscience and Remote Sensing Symposium (IGARSS)}, 
  title={Using Satellite-Derived SST Fronts to Evaluate an Eddy Resolving Numerical Circulation Model}, 
  year={2008},
  volume={2},
  pages={5-8}}

@article{COURTOIS201760,
author = {Peggy Courtois and Xianmin Hu and Clark Pennelly and Paul Spence and Paul G. Myers},
title = {Mixed layer depth calculation in deep convection regions in ocean numerical models},
journal = {Ocean Modelling},
volume = {120},
pages = {60-78},
year = {2017}}

@article{sr23review,
title = {Image super-resolution: A comprehensive review, recent trends, challenges and applications},
journal = {Information Fusion},
volume = {91},
pages = {230-260},
year = {2023},
issn = {1566-2535},
author = {Dawa Chyophel Lepcha and Bhawna Goyal and Ayush Dogra and Vishal Goyal}}

@article{yan22tmm,
  author={Yan, Yitong and Liu, Chuangchuang and Chen, Changyou and Sun, Xianfang and Jin, Longcun and Peng, Xinyi and Zhou, Xiang},
  journal={IEEE Transactions on Multimedia}, 
  title={Fine-Grained Attention and Feature-Sharing Generative Adversarial Networks for Single Image Super-Resolution}, 
  year={2022},
  volume={24},
  pages={1473-1487}}

@inproceedings{zhang18eccv,
author={Yulun, Zhang and Kunpeng, Li and Kai, Li and Lichen, Wang and Bineng, Zhong and Yun, Fu},
title = {Image super-resolution using very deep residual channel attention networks},
booktitle = {Proceedings of European Conference on Computer Vision (ECCV)},
year = {2018},
pages = {286-301}}

@inproceedings{swinir,
author={Liang, Jingyun and Cao, Jiezhang and Sun, Guolei and Zhang, Kai and Van Gool, Luc and Timofte, Radu},
booktitle={IEEE/CVF International Conference on Computer Vision Workshops (ICCVW)}, 
title={{SwinIR}: Image Restoration Using Swin Transformer}, 
year={2021},
pages={1833-1844}}

@inproceedings{zhou23iccv,
  author={Zhou, Yupeng and Li, Zhen and Guo, Chun-Le and Bai, Song and Cheng, Ming-Ming and Hou, Qibin},
  booktitle={Proceedings of IEEE/CVF International Conference on Computer Vision (ICCV)}, 
  title={{SRFormer}: Permuted Self-Attention for Single Image Super-Resolution}, 
  year={2023},
  pages={12734-12745}}

@inproceedings{yoo23wacv,
  author={Yoo, Jinsu and Kim, Taehoon and Lee, Sihaeng and Kim, Seung Hwan and Lee, Honglak and Kim, Tae Hyun},
  booktitle={Proceedings of IEEE/CVF Winter Conference on Applications of Computer Vision (WACV)}, 
  title={Enriched {CNN-Transformer} Feature Aggregation Networks for Super-Resolution}, 
  year={2023},
  pages={4945-4954}}

@misc{vim2024arxiv,
  title={Vision Mamba: Efficient Visual Representation Learning with Bidirectional State Space Model}, 
  author={Lianghui Zhu and Bencheng Liao and Qian Zhang and Xinlong Wang and Wenyu Liu and Xinggang Wang},
  year={2024},
  eprint={2401.09417},
  archivePrefix={arXiv},
  primaryClass={cs.CV}}

@Article{ali2023mdpi,
AUTHOR = {Ali, Anas M. and Benjdira, Bilel and Koubaa, Anis and El-Shafai, Walid and Khan, Zahid and Boulila, Wadii},
TITLE = {Vision Transformers in Image Restoration: A Survey},
JOURNAL = {Sensors},
VOLUME = {23},
YEAR = {2023},
NUMBER = {5},
ARTICLE-NUMBER = {2385},
URL = {https://www.mdpi.com/1424-8220/23/5/2385},
PubMedID = {36904589},
ISSN = {1424-8220},
DOI = {10.3390/s23052385}}

@article{8113128,
  author={Zhu, Xiao Xiang and Tuia, Devis and Mou, Lichao and Xia, Gui-Song and Zhang, Liangpei and Xu, Feng and Fraundorfer, Friedrich},
  journal={IEEE Geoscience and Remote Sensing Magazine}, 
  title={Deep Learning in Remote Sensing: A Comprehensive Review and List of Resources}, 
  year={2017},
  volume={5},
  number={4},
  pages={8-36},
  doi={10.1109/MGRS.2017.2762307}}

@article{Hutgrs,
  author={Hu, Ting and Zhang, Feng and Li, Wei and Hu, Weidong and Tao, Ran},
  journal={IEEE Transactions on Geoscience and Remote Sensing}, 
  title={Microwave Radiometer Data Superresolution Using Image Degradation and Residual Network}, 
  year={2019},
  volume={57},
  number={11},
  pages={8954-8967},
  doi={10.1109/TGRS.2019.2923886}}

@article{Richard,
  title={Daily High Resolution Blended Analyses for Sea Surface Temperature},
  author={Richard W. Reynolds and Thomas M. Smith and Chunying Liu and Dudley B. Chelton and Kenneth S. Casey and Michael G. Schlax},
  journal={American Meteorological Society},
  year={2007}}

@article{9714397,
  author={Liu, Xiaomin and Feng, Tiantian and Shen, Xiaofan and Li, Rongxing},
  journal={IEEE Transactions on Geoscience and Remote Sensing}, 
  title={PMDRnet: A Progressive Multiscale Deformable Residual Network for Multi-Image Super-Resolution of AMSR2 Arctic Sea Ice Images}, 
  year={2022},
  volume={60},
  pages={1-18}}

@inproceedings{srcnn,
  author={Dong, Chao and Loy, Chen Change and He, Kaiming and Tang, Xiaoou},
  booktitle={European Conference on Computer Vision (ECCV)}, 
  title={Learning a Deep Convolutional Network for Image Super-Resolution}, 
  year={2014},
  pages={184--199}}

@inproceedings{elan,
  title={Efficient Long-Range Attention Network for Image Super-Resolution},
  author={Zhang, Xindong and Zeng, Hui and Guo, Shi and Zhang, Lei},
  booktitle={Proceedings of European Conference on Computer Vision (ECCV)},
  year={2020},
  pages={649--667}}

@inproceedings{ducournau16prrs,
  author={Ducournau, Aurelien and Fablet, Ronan},
  booktitle={IAPR Workshop on Pattern Recogniton in Remote Sensing (PRRS)}, 
  title={Deep learning for ocean remote sensing: an application of convolutional neural networks for super-resolution on satellite-derived {SST} data}, 
  year={2016},
  pages={1-6}}

@INPROCEEDINGS{chazhi,
  author={ Chang, Hong and Yeung, Dit-Yan and Xiong, Yimin },
  booktitle={Proceedings of IEEE Conference on Computer Vision and Pattern Recognition (CVPR)},
  title={Super-resolution through neighbor embedding}, 
  year={2004},
  pages={1--8}}

@ARTICLE{chong,
  author={Farsiu, S. and Robinson, M.D. and Elad, M. and Milanfar, P.},
  journal={IEEE Transactions on Image Processing}, 
  title={Fast and robust multiframe super resolution}, 
  year={2004},
  volume={13},
  number={10},
  pages={1327-1344}}

@ARTICLE{dl,
  author={Zhu, Xiao Xiang and Tuia, Devis and Mou, Lichao and Xia, Gui-Song and Zhang, Liangpei and Xu, Feng and Fraundorfer, Friedrich},
  journal={IEEE Geoscience and Remote Sensing Magazine}, 
  title={Deep Learning in Remote Sensing: A Comprehensive Review and List of Resources}, 
  year={2017},
  volume={5},
  number={4},
  pages={8-36}}

@ARTICLE{ibp,
  author={Yoo, Jun-Sang and Kim, Jong-Ok},
  journal={IEEE Transactions on Image Processing}, 
  title={Noise-Robust Iterative Back-Projection}, 
  year={2020},
  volume={29},
  number={},
  pages={1219-1232},
  doi={10.1109/TIP.2019.2940414}}

@INPROCEEDINGS{spare,
  author={Bhuvaneswari, N. R. and Sivakumar, V. G.},
  booktitle={2016 International Conference on Communication and Electronics Systems (ICCES)}, 
  title={A comprehensive review on sparse representation for image classification in remote sensing}, 
  year={2016},
  volume={},
  number={},
  pages={1-4},
  doi={10.1109/CESYS.2016.7889923}}

@ARTICLE{dlm,
  author={Bordone Molini, Andrea and Valsesia, Diego and Fracastoro, Giulia and Magli, Enrico},
  journal={IEEE Transactions on Geoscience and Remote Sensing}, 
  title={DeepSUM: Deep Neural Network for Super-Resolution of Unregistered Multitemporal Images}, 
  year={2020},
  volume={58},
  number={5},
  pages={3644-3656},
  doi={10.1109/TGRS.2019.2959248}}

@Inproceedings{vdsr,
  author={Kim, Jiwon and Lee, Jung Kwon and Lee, Kyoung Mu},
  booktitle={Proceedings of IEEE Conference on Computer Vision and Pattern Recognition (CVPR)}, 
  title={Accurate Image Super-Resolution Using Very Deep Convolutional Networks}, 
  year={2016},
  pages={1646-1654}}

@article{ping21jstars,
  author={Ping, Bo and Su, Fenzhen and Han, Xingxing and Meng, Yunshan},
  journal={IEEE Journal of Selected Topics in Applied Earth Observations and Remote Sensing}, 
  title={Applications of Deep Learning-Based Super-Resolution for Sea Surface Temperature Reconstruction}, 
  year={2021},
  volume={14},
  pages={887-896}}

@article{su,
  title = {Super-resolution of subsurface temperature field from remote sensing observations based on machine learning},
  author = {Hua Su and An Wang and Tianyi Zhang and Tian Qin and Xiaoping Du and Xiao-Hai Yan},
  journal = {International Journal of Applied Earth Observation and Geoinformation},
  volume = {102},
  pages = {102440},
  year = {2021}}

@article{kim23ija,
  title = {Multi-source deep data fusion and super-resolution for downscaling sea surface temperature guided by Generative Adversarial Network-based spatiotemporal dependency learning},
  author = {Jinah Kim and Taekyung Kim and Joon-Gyu Ryu},
  journal = {International Journal of Applied Earth Observation and Geoinformation},
  volume = {119},
  pages = {103312},
  year = {2023}}

@article{zou23rs,
AUTHOR = {Zou, Runtai and Wei, Li and Guan, Lei},
TITLE = {Super Resolution of Satellite-Derived Sea Surface Temperature Using a Transformer-Based Model},
JOURNAL = {Remote Sensing},
VOLUME = {15},
YEAR = {2023},
NUMBER = {22},
pages={1-18}}

@article{gu21ssm,
  title={Efficiently Modeling Long Sequences with Structured State Spaces}, 
  author={Albert Gu and Karan Goel and Christopher Ré},
  year={2021},
  journal={arXiv preprint arXiv:2111.00396}}

@inproceedings{mehta22long,
  title={Long Range Language Modeling via Gated State Spaces}, 
  author={Harsh Mehta and Ankit Gupta and Ashok Cutkosky and Behnam Neyshabur},
  year={2023},
  booktitle={Proceedings of International Conference on Learning Representations (ICLR)},
  pages={1-13}}

@article{gu23mamba,
title={Mamba: Linear-Time Sequence Modeling with Selective State Spaces}, 
author={Albert Gu and Tri Dao},
year={2023},
journal={arXiv preprint arXiv:2312.00752}}

@article{liu24cmunet,
  title={{CM-UNet}: Hybrid {CNN-Mamba UNet} for Remote Sensing Image Semantic Segmentation}, 
  author={Mushui Liu and Jun Dan and Ziqian Lu and Yunlong Yu and Yingming Li and Xi Li},
  year={2024},
  journal={arXiv preprint arXiv:2405.10530}}

@article{zhou24rsde,
  title={{RSDehamba}: Lightweight Vision {Mamba} for Remote Sensing Satellite Image Dehazing}, 
  author={Huiling Zhou and Xianhao Wu and Hongming Chen and Xiang Chen and Xin He},
  year={2024},
  journal={arXiv preprint arXiv:2405.10030}}

@Article{rs13183568,
  AUTHOR = {Tomoki， Izumi and Motoki， Amagasaki and Kei， Ishidaand and Masato， Kiyama},
  TITLE = {Super-resolution of sea surface temperature with convolutional neural network- and generative adversarial network-based methods},
  JOURNAL = {Journal of water and climate change},
  VOLUME = {13},
  YEAR = {2022},
  issue = {4},
  pages = {1673-1683}}

@article{vgg19,
  title={Very Deep Convolutional Networks for Large-Scale Image Recognition},
  author={Karen Simonyan and Andrew Zisserman},
  journal={CoRR},
  year={2014},
  volume={},
  url={https://api.semanticscholar.org/CorpusID:14124313}
}

@ARTICLE{dwt,
  author={Liu, Pengju and Zhang, Hongzhi and Lian, Wei and Zuo, Wangmeng},
  journal={IEEE Access}, 
  title={Multi-Level Wavelet Convolutional Neural Networks}, 
  year={2019},
  volume={7},
  number={},
  pages={74973-74985},
  doi={10.1109/ACCESS.2019.2921451}}

@INPROCEEDINGS{ffl,
  author={Jiang, Liming and Dai, Bo and Wu, Wayne and Loy, Chen Change},
  booktitle={2021 IEEE/CVF International Conference on Computer Vision (ICCV)}, 
  title={Focal Frequency Loss for Image Reconstruction and Synthesis}, 
  year={2021},
  volume={},
  number={},
  pages={13899-13909},
  doi={10.1109/ICCV48922.2021.01366}}

@INPROCEEDINGS{fpl,
  author={Fuoli, Dario and Van Gool, Luc and Timofte, Radu},
  booktitle={2021 IEEE/CVF International Conference on Computer Vision (ICCV)}, 
  title={Fourier Space Losses for Efficient Perceptual Image Super-Resolution}, 
  year={2021},
  volume={},
  number={},
  pages={2340-2349},
  doi={10.1109/ICCV48922.2021.00236}}

@electronic{hycom,
    url     =   {{https://www.hycom.org/dataserver/gofs-3pt1/reanalysis}},
    title   =   {{GOFS 3.1: 41-layer HYCOM + NCODA Global 1/12° Reanalysis}}
}

@electronic{ghrsst,
    url     =   {{https://www.ghrsst.org/ghrsst-data-services/for-sst-data-users}},
    title   =   {{FOR SST DATA USERS}}
}

@electronic{oisst,
    url     =   {{https://www.ncei.noaa.gov/products/optimum-interpolation-sst}},
    title   =   {{Optimum Interpolation SST}}
}
\bibliographystyle{IEEEtran}

\end{document}